\documentclass[sigplan,screen]{acmart}
\AtBeginDocument{%
  \providecommand\BibTeX{{%
    \normalfont B\kern-0.5em{\scshape i\kern-0.25em b}\kern-0.8em\TeX}}}

\acmYear{2022}\copyrightyear{2022}
\setcopyright{acmcopyright}
\acmConference[PPoPP '22]{27th ACM SIGPLAN Symposium on Principles and Practice of Parallel Programming}{April 2--6, 2022}{Seoul, Republic of Korea}
\acmBooktitle{27th ACM SIGPLAN Symposium on Principles and Practice of Parallel Programming (PPoPP '22), April 2--6, 2022, Seoul, Republic of Korea}
\acmPrice{15.00}
\acmDOI{10.1145/3503221.3508409}
\acmISBN{978-1-4503-9204-4/22/04}

\usepackage{xspace}
\usepackage{subfig}
\usepackage[compact]{titlesec}

\usepackage{flushend}
\usepackage{color}
\definecolor{lightgray}{rgb}{0.95, 0.95, 0.95}
\definecolor{darkgray}{rgb}{0.4, 0.4, 0.4}
\definecolor{editorGray}{rgb}{0.95, 0.95, 0.95}
\definecolor{editorOcher}{rgb}{1, 0.5, 0} 
\definecolor{editorGreen}{rgb}{0, 0.5, 0} 
\definecolor{orange}{rgb}{1,0.45,0.13}
\definecolor{olive}{rgb}{0.17,0.59,0.20}
\definecolor{brown}{rgb}{0.69,0.31,0.31}
\definecolor{purple}{rgb}{0.38,0.18,0.81}
\definecolor{lightblue}{rgb}{0.1,0.57,0.7}
\definecolor{lightred}{rgb}{1,0.4,0.5}
\usepackage{upquote}
\usepackage{listings}

\lstdefinelanguage{mylang}{
  keywords={if, else, foreach, in, return},
  ndkeywords={input, RG_Shader, IS_Shader, AH_Shader, buildBVH, traceRays, reorderQueries},
  sensitive=false,
  comment=[l]{//},
  morecomment=[s]{/*}{*/},
  morestring=[b]',
  morestring=[b]"
}

\titlespacing*{\section}{0pt}{8pt plus 0pt minus 0pt}{4pt plus 0pt minus 0pt}
\titlespacing*{\subsection}{0pt}{6pt plus 0pt minus 0pt}{3pt plus 0pt minus 0pt}

\ifx\figurename\undefined \def\figurename{Figure}\fi
\renewcommand{\figurename}{Fig.}
\renewcommand{\paragraph}[1]{\textbf{#1} }

\newcommand{\Sec}[1]{Section~\ref{#1}}
\newcommand{\Fig}[1]{Figure~\ref{#1}}

\newcommand{\Lst}[1]{Listing~\ref{#1}}

\newcommand{\PBox}[1]{\vspace*{.05cm}\noindent\fbox{\parbox{\columnwidth}{\vspace*{.05cm}{#1}}}\vspace*{.05cm}}

\newcommand{\proj}{\textsc{RTNN}\xspace}

\newcommand{\is}{\textsf{IS}\xspace}
\newcommand{\ah}{\textsf{AH}\xspace}
\newcommand{\rg}{\textsf{RG}\xspace}

\newcommand{\no}[1]{#1}
\renewcommand{\no}[1]{}
\newcommand{\RNum}[1]{\uppercase\expandafter{\romannumeral #1\relax}}

\graphicspath{{figs/}}

\begin{document}

\title{RTNN: Accelerating Neighbor Search Using Hardware Ray Tracing}

\author{Yuhao Zhu}
\email{yzhu@rochester.edu}
\affiliation{%
  \institution{University of Rochester}
  \city{Rochester}
  \state{New York}
  \country{USA}
}


\begin{abstract}
Neighbor search is of fundamental importance to many engineering and science fields such as physics simulation and computer graphics. This paper proposes to formulate neighbor search as a ray tracing problem and leverage the dedicated ray tracing hardware in recent GPUs for acceleration. We show that a naive mapping under-exploits the ray tracing hardware. We propose two performance optimizations, query scheduling and query partitioning, to tame the inefficiencies. Experimental results show 2.2$\times$ -- 65.0$\times$ speedups over existing neighbor search libraries on GPUs. The code is available at \url{https://github.com/horizon-research/rtnn}.
\end{abstract}



\maketitle

\section{Introduction}
\label{sec:intro}

3D Neighbor search is a building block widely used in many application domains such as computer vision, graphics, and scientific computing. Due to its fundamental importance, fast neighbor search has long been a subject of much research, including many CPU~\cite{ihmsen2011parallel, muja2009fast, CompactNSearch} and GPU libraries~\cite{hoetzlein2014fast, cunsearchcode, frnn} as well as hardware accelerators~\cite{xu2019tigris, pinkham2020quicknn}.

A fundamental trade-off neighbor search algorithms make is one between work efficiency and hardware efficiency. On one hand, grid-based algorithms are work-inefficient, as they perform (limited) exhaustive search over a grid, but exhaustive searches are hardware friendly and can be easily parallelized. On the other hand, tree-based algorithms (e.g., Octree, k-d tree) are work-efficient by hierarchically pruning the search space, but tree traversal is hardware-inefficient, introducing irregular control flows and memory accesses.


This paper argues that neighbor search can be made both work-efficient and hardware-friendly --- by using Bounding Volume Hierarchy (BVH) tree as the basic data structure. This design decision allows us to formulate neighbor search as a ray tracing problem (\Sec{sec:idea:basic}), which, critically, has dedicated hardware support (for BVH traversal) in recent GPUs such as Nvidia's Turing (and later) architecture.





Unfortunately, a naive mapping from neighbor search to ray tracing does not effectively exploit the ray tracing hardware, and is in fact work- and hardware-inefficient (\Sec{sec:idea:char}). We quantitatively show two performance-limiting factors : 1) unmanaged query-to-ray mapping, which leads to control-flow divergences, and 2) excessive tree traversals stemming from the monolithic BVH construction. These characterizations motivate us to propose two optimizations.


First, we propose a query scheduling strategy to tame the control-flow divergence by mapping spatially-close queries to nearby rays (\Sec{sec:sched}). We show that this scheduling algorithm, in itself, can be formulated as a truncated ray tracing problem and, thus, is extremely efficient to execute.

Second, we propose a lightweight query partitioning algorithm to aggressively suppress tree traversals (\Sec{sec:part:idea}). Instead of using a single BVH for all the queries/rays, we partition queries such that each partition has a unique BVH that minimizes tree traversals for that partition. Query partitioning, however, comes with the overhead of extra BVH constructions. We propose an algorithm that bundles the partitions to minimize the execution time (\Sec{sec:part:opt}).


On the RTX 2080 GPU, we show 2.2$\times$ to 44.0$\times$ speedups compared to optimized CUDA neighbor search and a 65.0$\times$ speedup over unoptimized ray tracing-accelerated neighbor search. The contributions of the paper are the following:

\begin{figure*}[t]
    \centering
    \subfloat[A simple scene (left) with three primitives (triangles here) and the corresponding BVH (right). Numbers denote AABBs and primitives. Each BVH node represents an AABB, and the leaf nodes store the actual primitives. Each leaf node (3, 5, 7) in this example stores one primitive, but in principle more primitives per leaf node is possible.]{
      \label{fig:bvh-example}
      \includegraphics[height=1.1in]{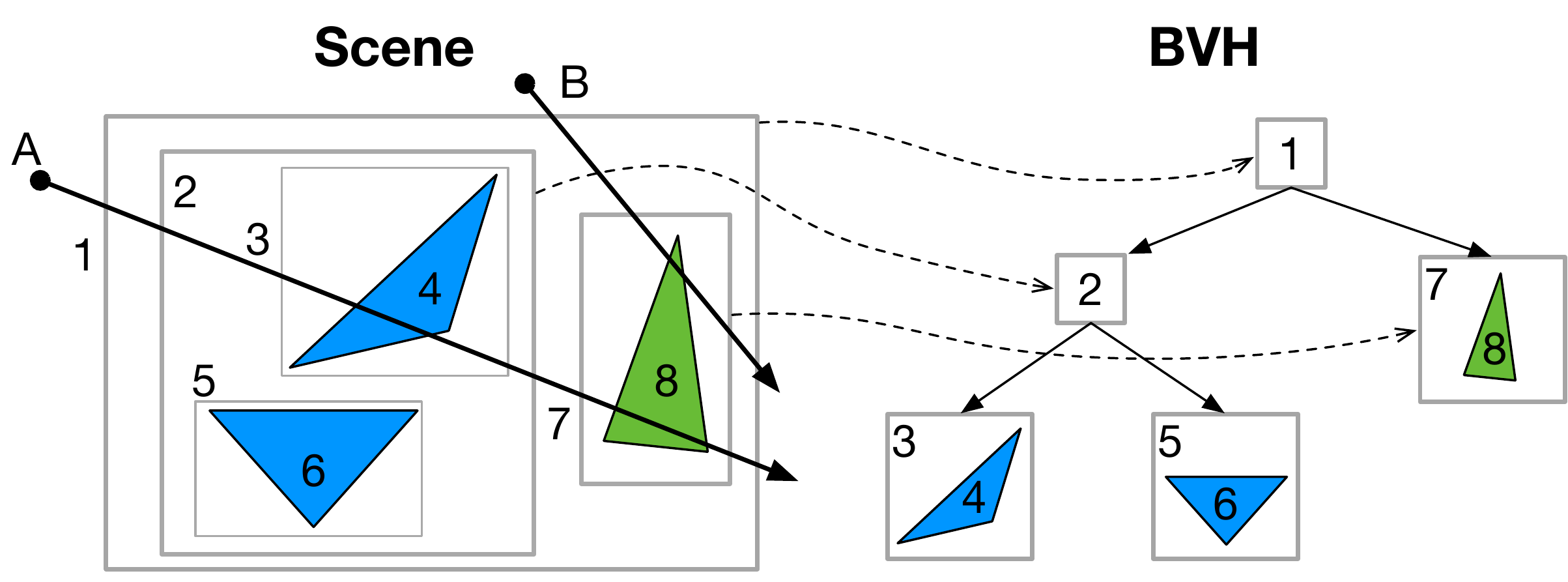}
    }
    \hspace{5pt}
    \subfloat[The (not-to-scale) timelines of tracing Ray A and Ray B on the BVH. TL represents traversal (including ray-AABB intersection test), which executes on the RT cores. Ray A and Ray B are spatially apart; they exercise different traversal paths and shaders. \textsf{IS} shader is skipped for primitives whose AABBs do not intersect the ray. No \textsf{AH}/\textsf{Miss} shader in this example.]{
      \label{fig:optix-timeline}
      \includegraphics[height=1.1in]{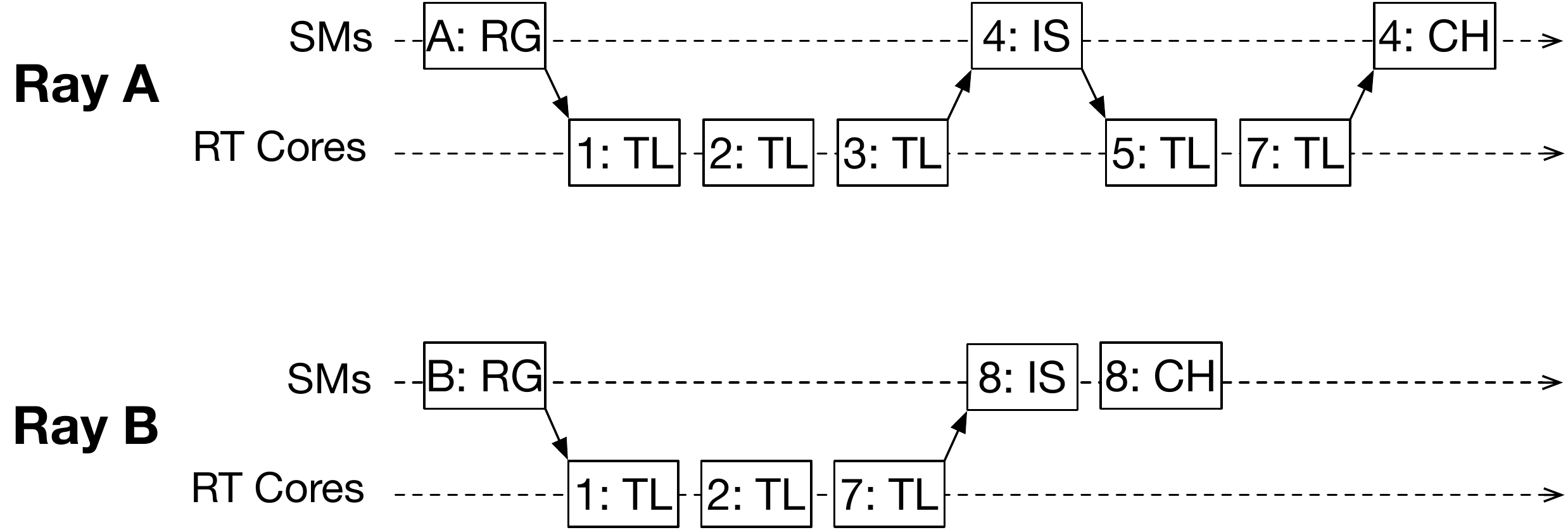}
    }
    \caption{A simple BVH example and the (not-to-scale) execution timelines of tracing two rays on the BVH. Abbreviations: RG (Ray Generation), TL (Traversal), IS (Intersection), CH (Closest-Hit), AH (Any-Hit); see \Fig{fig:optix-progmodel}.}
    \label{fig:module}
\end{figure*}

\begin{itemize}
	\item We describe a systematic way to map neighbor search to ray tracing, and quantitatively demonstrate two key performance bottlenecks of such a mapping.
	\item We introduce two optimizations, query scheduling and query partitioning, that mitigate the bottlenecks and effectively exploit the ray tracing hardware.
	\item We provide an open-source implementation of our algorithm, which achieves 2.2$\times$ -- 65.0$\times$ speedup over existing GPU neighbor search algorithms.
\end{itemize}

\section{Background}
\label{sec:bck}

We first define the scope of neighbor search that is considered in this paper (\Sec{sec:bg:ns}). We then briefly overview the ray tracing algorithm (\Sec{sec:bg:algo}) pertaining to this paper, and introduce the programming and hardware support for ray tracing in Nvidia's recent GPUs (\Sec{sec:bg:hw}).

\subsection{Scope of Neighbor Search This Paper Targets}
\label{sec:bg:ns}

\paragraph{Dimensionality in Neighbor Search} Different applications require neighbor searches in different dimensions. Due to the curse of dimensionality, it is well-known that search algorithms used for low dimensions (three or lower) are different from that for high-dimensional searches~\cite{beyer1999nearest, weber1998quantitative, walters2010comparative}.


We target neighbor search in low-dimensional (three or lower) space, which is prevalent in engineering and science fields (e.g., computational fluid dynamics, graphics, vision), because they deal with physical data such as particles and surface samples that inherently reside in the 2D/3D space.



\paragraph{Neighbor Search Variants} Two types of neighbor search exist: fixed-radius search (a.k.a., range search) and K nearest neighbor search. \proj optimizes for both types.

Fixed-radius search concerns with returning all the neighbors within a fixed radius $r$. In practice, the maximum amount of returned neighbors is bounded in order to bound the memory consumption and to interface with downstream tasks, which usually expect a fixed amount of neighbors.


KNN search concerns with returning the nearest K neighbors of a query. In practice, the returned neighbors are bounded by a search radius, beyond which the neighbors are discarded. This is because the significance of a neighbor (e.g., the force that a particle exerts on another) is minimal and of little interest when it is too far away.

Therefore, for both types of search we assume a search interface that provides a search radius $r$ and a maximum neighbor count $K$, consistent with the interface of existing neighbor search libraries. We can easily emulate an unbounded KNN search by providing a very large $r$ and emulate an unbounded range search by providing a very large $K$.



\subsection{Ray Tracing Algorithm and Data Structure}
\label{sec:bg:algo}

Graphics rendering algorithms are moving toward ray tracing. We briefly review algorithmic components relevant to our paper, and refer interested readers to Pharr et al.~\cite{pharr2016physically} and Glassner~\cite{glassner1989introduction} for a more comprehensive treatment.

\paragraph{Intersection Test} The crux of ray tracing is to calculate the closest intersection of a ray and the scene, which is usually represented by a set of geometric primitives such as triangles and spheres. The intersection test dominates the rendering time~\cite{vasiou2018detailed}, and is the prime target for optimization.



The intersection test is done by partitioning the primitives in the scene. In particular, primitives are represented by their bounding volumes, which are usually Axis-Aligned Bounding Boxes (AABBs). The AABBs are then hierarchically organized as a tree, which is called the Bounding Volume Hierarchy (BVH). \Fig{fig:bvh-example} shows the BVH of a simple three-primitive (triangles here) scene. The leaf nodes in the BVH are the AABBs that store the actual scene primitives, and the interior nodes are the AABBs that enclose other AABBs.

With the BVH, finding the closest hit for a ray becomes a tree traversal problem. At every node, we test whether the ray intersects with the AABB of that node. If the ray does not intersect the node's AABB, the entire subtree beneath that node can be skipped, because all the primitives enclosed by that AABB are guaranteed to be not intersected by the ray. For instance, Ray A in \Fig{fig:bvh-example} does not intersect AABB 5, so primitive 6 can be skipped. Otherwise, we further test the ray against all the AABBs enclosed by the node. Note that AABB 7 and primitive 8 will also be skipped after hit test with primitive 4, which provides a closer hit than AABB 7.

When the ray reaches a leaf node, we test the ray against all the enclosed primitives and record the closet hit point (so far). This process continues until we have traversed the entire tree, at which point the closet hit is reported.

\paragraph{Intersection Conditions} It is vital to understand the conditions under which an AABB is considered to be intersected by a ray, which our algorithm relies on. Formally, a ray is a line $P(t)$ parameterized by two parameters: the origin $O$ and the direction vector $\mathbf{d}$~\cite{Shirley2019}:
\begin{align}
	P(t) = O + t\mathbf{d}.
\end{align}

\begin{figure}[h]
\centering
\includegraphics[width=\columnwidth]{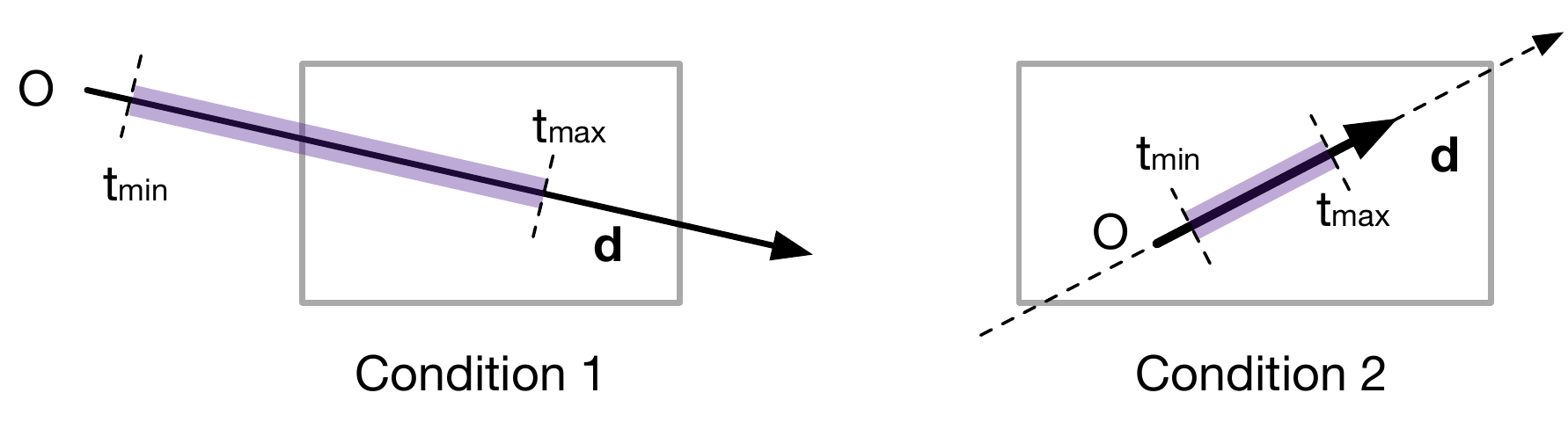}
\caption{Two conditions for ray-AABB intersection.}
\label{fig:intersect-cond}
\end{figure}

While in theory $t$ can take any value, providing a full line, in practice we are often interested in only a segment of the ray, which is described by bounding $t \in [t_{min}, t_{max}]$. A ray intersects an AABB if one of the following two conditions is met. \Fig{fig:intersect-cond} illustrates the two conditions.

\begin{enumerate}
	\item when a ray hits the bounds of the AABB (the six faces) and the $t$ value of hit point is within $[t_{min}, t_{max}]$;
	\item when the origin of the ray is within the AABB, \textit{even if} the intersected $t$ value is beyond $[t_{min}, t_{max}]$.
\end{enumerate}

As we will show later, we rely on Condition 2 to implement neighbor search. Condition 2 might initially seem odd. It is necessary because when a ray originates from within an AABB, it is possible that the ray \textit{might} intersect children AABBs that are enclosed in the current AABB. Therefore, we must treat that ray as intersecting such that the ray is allowed to further test against the enclosed (children) AABBs.

\subsection{Hardware Support and Programming Model for Ray Tracing on Nvidia GPUs}
\label{sec:bg:hw}

While using BVH to prune the search space is work-efficient, tree traversal is irregular, exhibiting frequent control-flow and memory divergences (e.g., per-thread stack management). Nvidia's recent Turing (and later) GPU architecture is equipped with dedicated hardware, i.e., the RT cores, to accelerate BVH tree traversal~\cite{burgess2020rtx}. We briefly review the architectural and programming details that are relevant to developing neighbor search algorithms.

\paragraph{Hardware} The RT cores are essentially tightly-coupled accelerators sitting alongside the conventional Stream Multiprocessors (SMs). The RT cores and the SMs share the same device memory --- an important feature that allows us to, in one program, use SMs for regular parallel computations and use the RT cores for ray tracing.

\begin{figure}[t]
\centering
\includegraphics[width=\columnwidth]{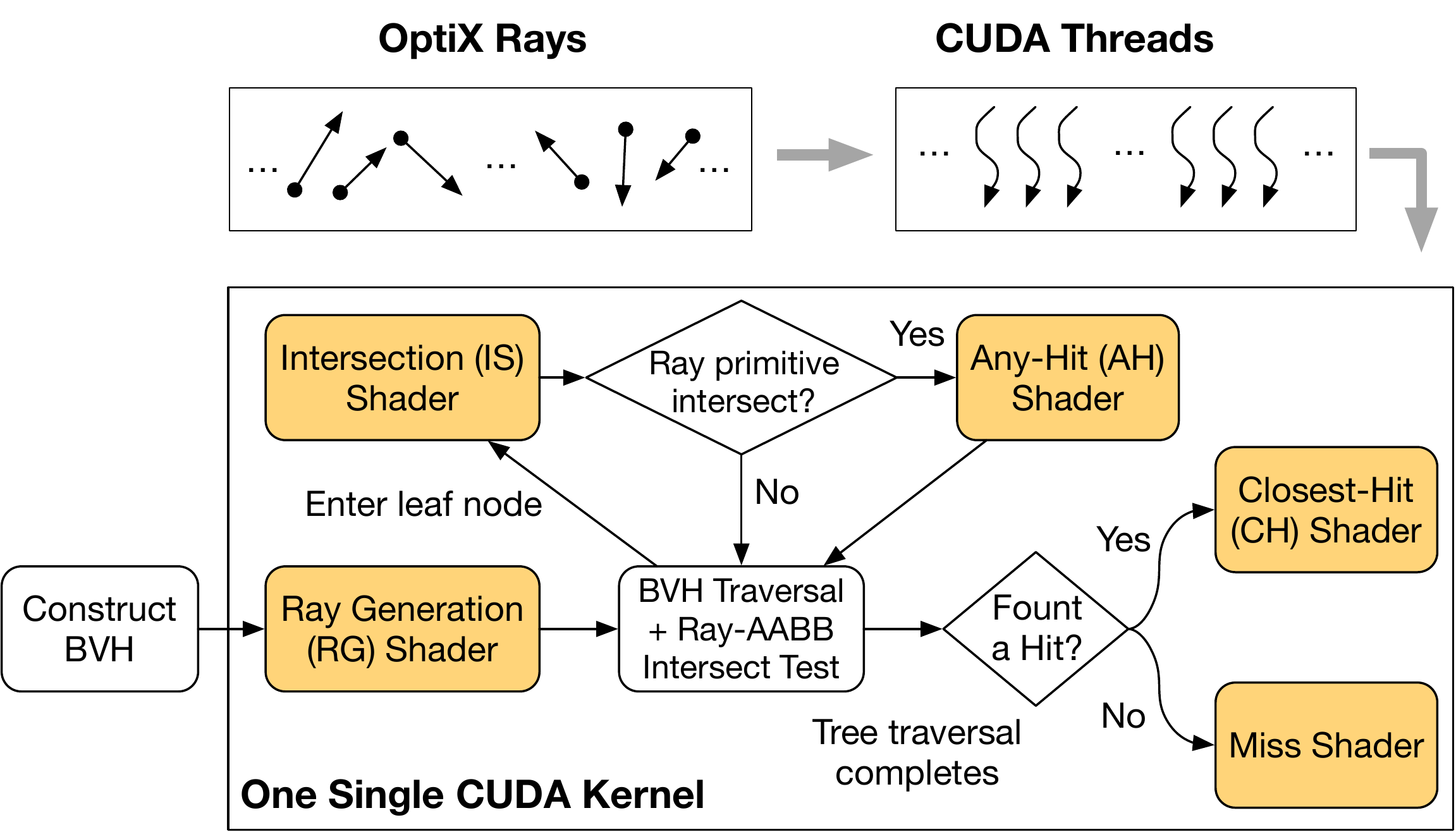}
\caption{The simplified programming model of OptiX. Shaded components are the programmable shaders. \textsf{AH}/\textsf{CH}/\textsf{Miss} shaders are optional. All the shaders are compiled into one single CUDA kernel, which executes on the SMs. Each ray is mapped to a CUDA thread. BVH traversals, including the ray-AABB intersection tests, are accelerated on the RT cores.}
\label{fig:optix-progmodel}
\end{figure}

\paragraph{Programming Model} To leverage the RT cores, we use the OptiX programming model~\cite{parker2010optix} from Nvidia as it natively supports the RT cores, but the same principles apply to other graphics APIs as well (e.g., Vulkan and DirectX).

In OptiX, ray tracing starts by building the BVH; this stage executes on the SMs and is non-programmable. Once the BVH is built, the ray tracing pipeline is launched. OptiX presents to programmers a fixed pipeline organization, but exposes interfaces for user-defined programs (a.k.a., \textit{shaders} in the graphics parlance) to control different stages in the pipeline. It is these programmable shaders that provide the opportunity for implementing algorithms beyond rendering. \Fig{fig:optix-progmodel} shows a simplified view of the pipeline, where the shaded components are the programmable shaders.

OptiX shaders are essentially callback functions triggered at different phases during BVH traversal. The Ray Generation (\textsf{RG}) shader, the entry to the pipeline, generates rays by specifying the ray origins and directions. During traversal, whenever leaf nodes of the BVH are encountered the Intersection (\textsf{IS}) shader is called, which performs the ray-primitive intersection test. If an intersection is found, the Any-Hit (\textsf{AH}) shader can be called to process the hit information or to terminate the traversal. When the entire traversal finishes, either the Closest-Hit (\textsf{CH}) shader or the \textsf{Miss} shader could be called depending on whether a hit is found.

\paragraph{Execution Model} OptiX provides a ``Single Instruction Multiple Rays'' execution model: every shader in the pipeline (\Fig{fig:optix-progmodel}) is executed by every single ray. Under the hood, all the shaders are compiled into one single CUDA kernel executing on the SMs; each ray is mapped to a CUDA thread.

\Fig{fig:optix-timeline} illustrates the execution timelines of tracing the two rays in the BVH in \Fig{fig:bvh-example}. Each ray starts from the \textsf{RG} shader on the SMs. The control then transfers to the RT cores for the BVH traversal, during which if a shader is triggered the hardware traversal is interrupted and the control is transferred back to the SMs.

\paragraph{Terminology Clarification} In graphics parlance, identifying the intersection of a ray and the scene is called \textit{ray casting}; ray tracing refers to \textit{recursive} ray casting; the recursion is necessary for realistic shading~\cite{kajiya1986rendering, whitted1979improved}. In Nvidia's post-Turing GPUs, it is ray casting that is being accelerated in hardware. How (and whether) to implement recursion is left to the OptiX programmers (usually in the \textsf{IS} shader). In this sense, our paper maps neighbor search to a ray casting problem rather than a ray tracing problem. That is, a query will not recursively spawn new queries.

\section{Neighbor Search as Ray Tracing: Basic Idea and Performance Characterizations}
\label{sec:idea}

We describe how to formulate neighbor search as ray tracing (\Sec{sec:idea:basic}). We then quantitatively demonstrate two sources that dictate the performance of our algorithm (\Sec{sec:idea:char}), i.e., ray coherence and the AABB size, which motivate our optimizations that follow.

\subsection{The Basic Idea}
\label{sec:idea:basic}

\begin{figure}[t]
    \centering
    \subfloat[Original neighbor search for query $Q$.]{
      \label{fig:rnn-orig}
      \includegraphics[height=1.27in]{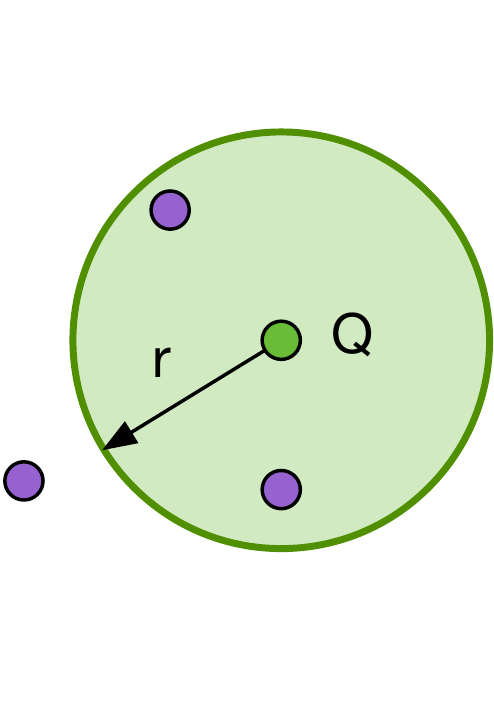}
    }
    \hfill
    \subfloat[Reversed neighbor search.]{
      \label{fig:rnn-new}
      \includegraphics[height=1.27in]{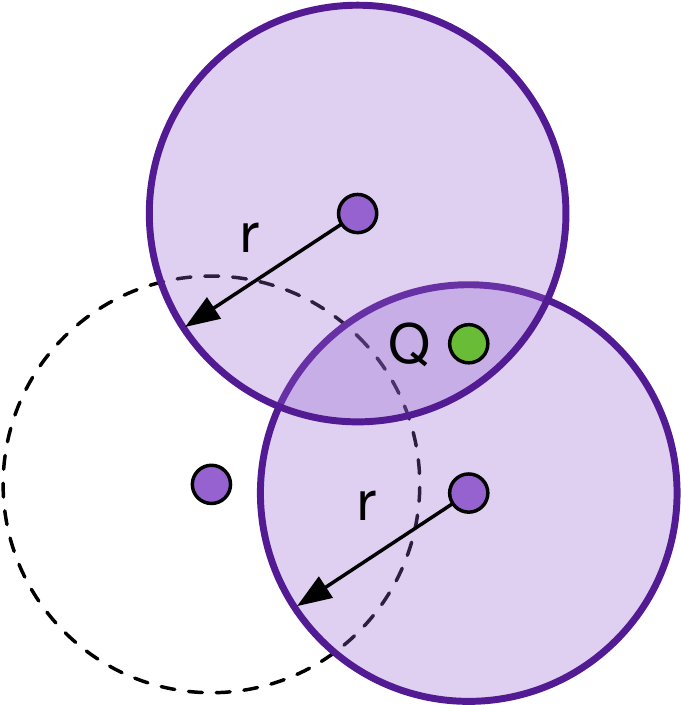}
    }
    \hfill
    \subfloat[Testing if $Q$ is in $P$'s $r$-sphere.]{
      \label{fig:rnn-test}
      \includegraphics[height=1.22in]{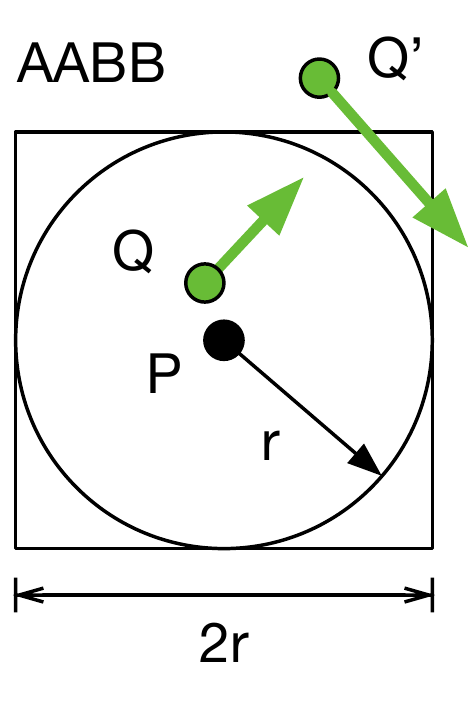}
    }
    \caption{Formulating neighbor search as ray tracing. (a): Searching points that are within $r$ from the query point $Q$; (b) Testing whether $Q$ is within $r$ from all other points; (c) Testing whether $Q$ is in $P$'s $r$-sphere can be done by tracing a very short ray from $Q$. Using long rays would lead to false positives in ray-AABB tests (e.g., $Q'$).}
    \label{fig:rnn-formulation}
\end{figure}

Distance measure in Euclidean spaces is commutative: testing whether a point $P$ is within a distance $r$ from a query $Q$ is equivalent to testing when $Q$ is within the same distance $r$ from $P$. Leveraging this property, we can turn the neighbor search problem around: instead of finding all the points within a distance $r$ from a query point $Q$ (shown in \Fig{fig:rnn-orig}), we test whether $Q$ is within $r$ from all other points. This inverse test can be done by generating spheres with a radius $r$ around all the points, and returning points whose spheres enclose $Q$ (shown in \Fig{fig:rnn-new}).

Identifying $r$-radius spheres that enclose $Q$ is done in two steps. \Fig{fig:rnn-test} illustrates the idea.

\begin{itemize}
	\item Step 1 (AABB test): For each sphere, we first generate an AABB that circumscribes the sphere (i.e., the tightest AABB that just encloses the sphere), and test whether $Q$ resides in the AABB. All the spheres that fail this test can be skipped.
	\item Step 2 (Sphere test): Otherwise, we further test whether $Q$ resides in the sphere, by comparing the distance between $Q$ and the sphere center $P$ with $r$. If successful, we can record $P$ as a neighbor of $Q$ under range search or operate a priority queue under KNN search.
\end{itemize}

Critically, the algorithm above can be mapped as a ray tracing problem. This is done by building a BVH from the AABBs of all the points and casting a ray originated from $Q$. Step 1 essentially uses the ray to traverse the BVH and discards points whose AABBs do not intersect the ray. Step 2 is a ray-primitive intersection test, where the primitives are spheres. Step 1 is completely accelerated by the RT cores, and Step 2 can be implemented as an \textsf{IS} shader in OptiX.

\paragraph{Casting the Ray} In theory, the ray from $Q$ can be of an arbitrary length. However, using a ray with a long length could lead to false positives in Step 1. \Fig{fig:rnn-test} illustrates this scenario using query $Q'$, whose ray intersects with $P$'s AABB (i.e., Step 1 test passes and Step 2 test is performed), but does not reside in $P$'s sphere.

While this false positive does not affect the correctness, as Step 2 will eventually reject $P$ as a neighbor of $Q$', it does lead to redundant computation in Step 2, which is much more expensive than Step 1 --- an order of magnitude slower in our experiments: Step 2 requires floating point multiplications and potentially manipulates a priority queue, whereas Step 1 requires only bounds comparison.

Therefore, we generate very short rays from the queries by setting $t_{min}$ to be 0 and $t_{max}$ to be a small number (e.g., 1e-16 in our implementation). With this, only rays whose origins reside in an AABB will trigger Step 2. Note that the ray-AABB intersection tests now will mostly rely on Condition 2 (\Fig{fig:intersect-cond}) since the rays are very short. With short rays, the ray direction can be arbitrary. We set all ray directions to  [1, 0, 0] in our implementation.

\paragraph{Benefits} Our algorithm is work-efficient, because it prunes the search space by omitting points whose AABBs do not contain the query point. It is similar, in spirit, to other tree-based algorithms using k-d trees~\cite{zhou2008real, flann} and Octrees~\cite{behley2015efficient, pcloctreeex}. Using BVH, however, let us leverage the ray tracing hardware in recent GPUs to accelerate the irregular tree traversals, which would not be available if other data structures were used.


\begin{figure*}[t]
\centering
\begin{minipage}[t]{0.48\columnwidth}
  \centering
  \includegraphics[trim=0 0 0 0, clip, height=1.15in]{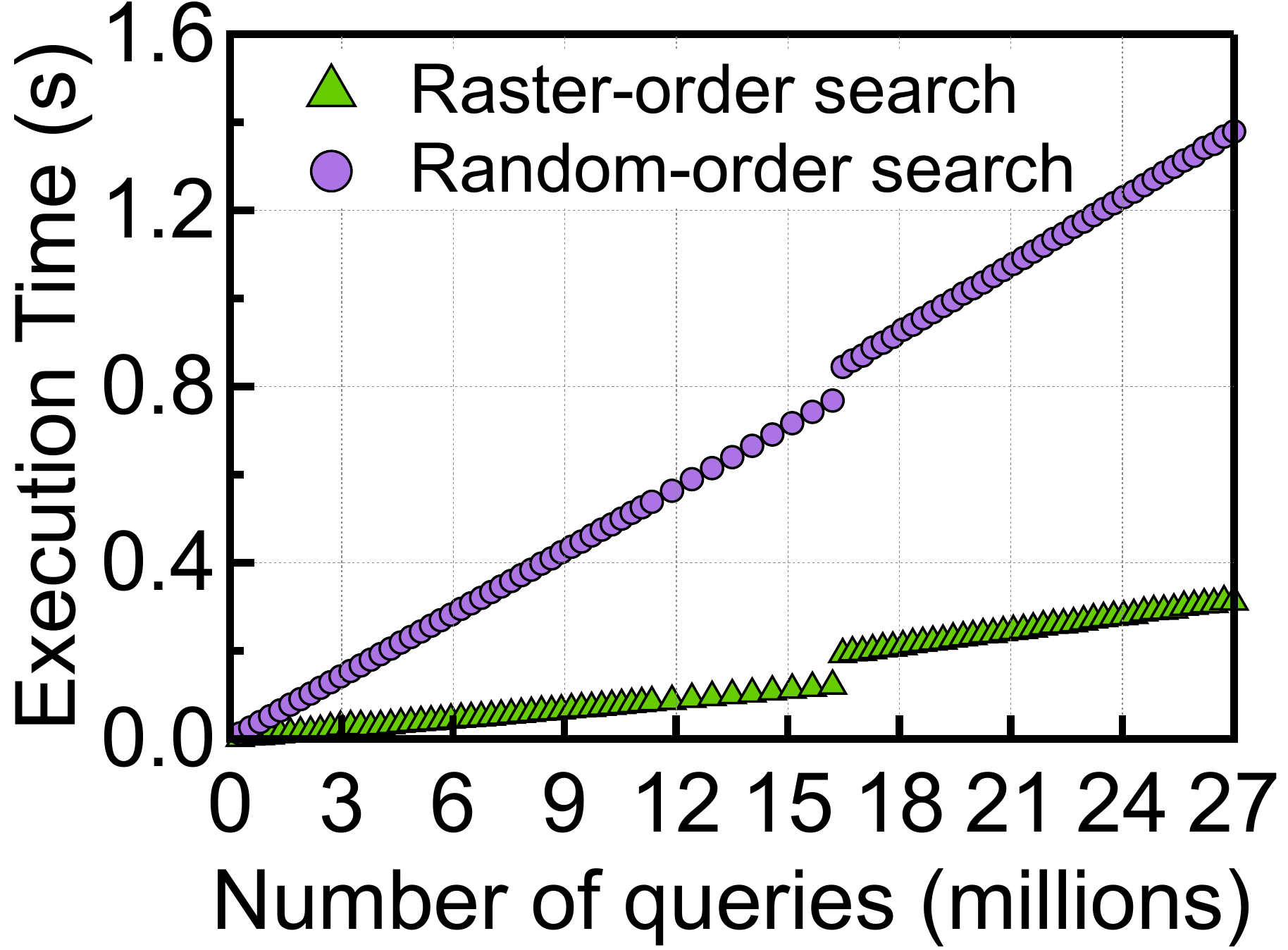}
  \caption{Searches with ordered (coherent) rays (raster-scan order here) are 4$\times$ faster than searches with incoherent, randomly-ordered rays.}
  \label{fig:random-vs-raster}
\end{minipage}
\hfill
\begin{minipage}[t]{0.48\columnwidth}
  \centering
  \includegraphics[trim=0 0 0 0, clip, height=1.15in]{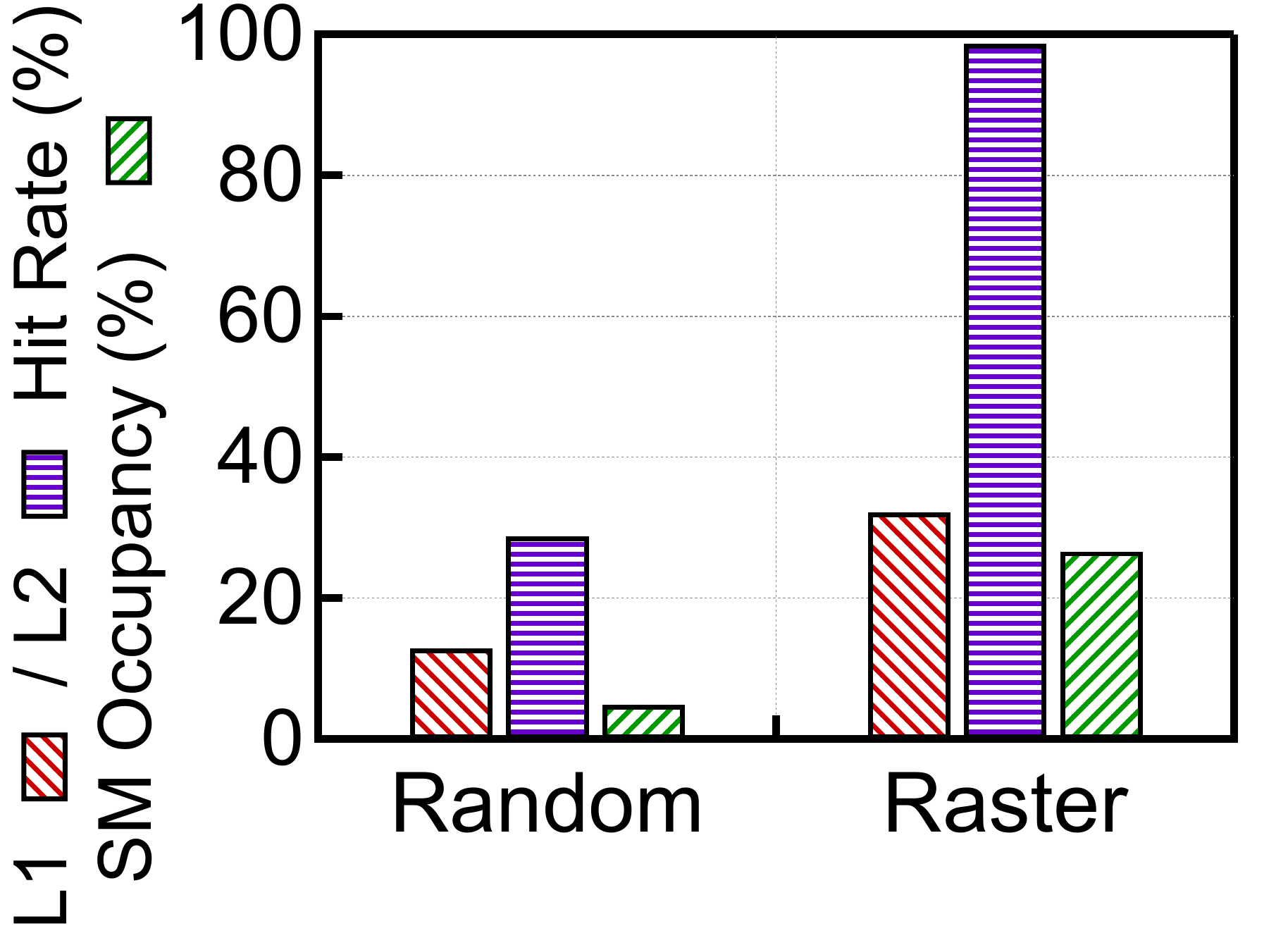}
  \caption{Ordered searches are faster because coherent queries/rays lead to higher L1/L2 cache hit rate and higher SM occupancy.}
  \label{fig:random-vs-raster-stat}
\end{minipage}
\hfill
\begin{minipage}[t]{0.48\columnwidth}
  \centering
  \includegraphics[trim=0 0 0 0, clip, height=1.15in]{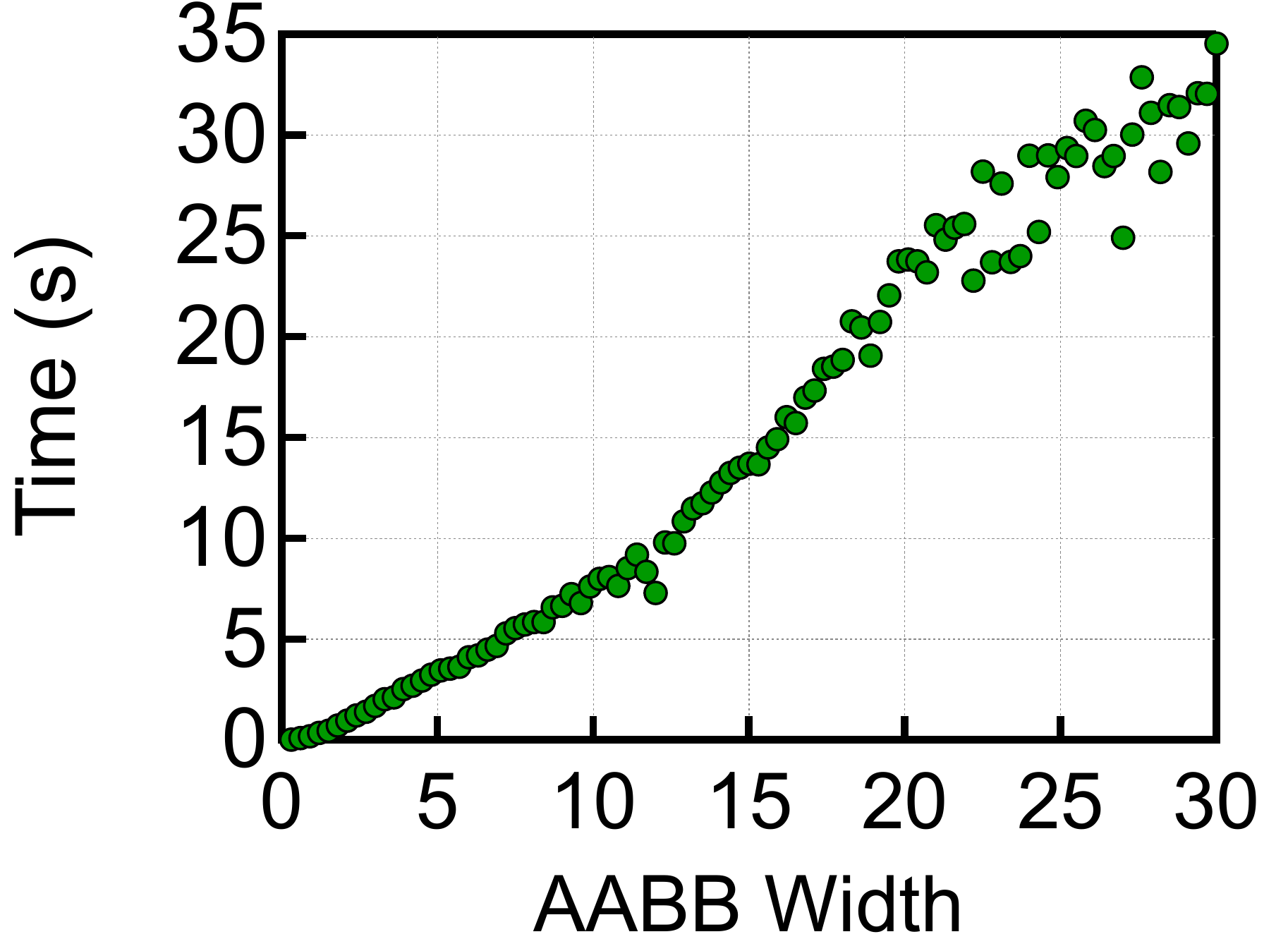}
  \caption{Search time correlates with the AABB size, which dictates the work on both the SMs and the RT cores.}
  \label{fig:time-vs-aabb}
\end{minipage}
\hfill
\begin{minipage}[t]{0.48\columnwidth}
  \centering
  \includegraphics[trim=0 0 0 0, clip, height=1.15in]{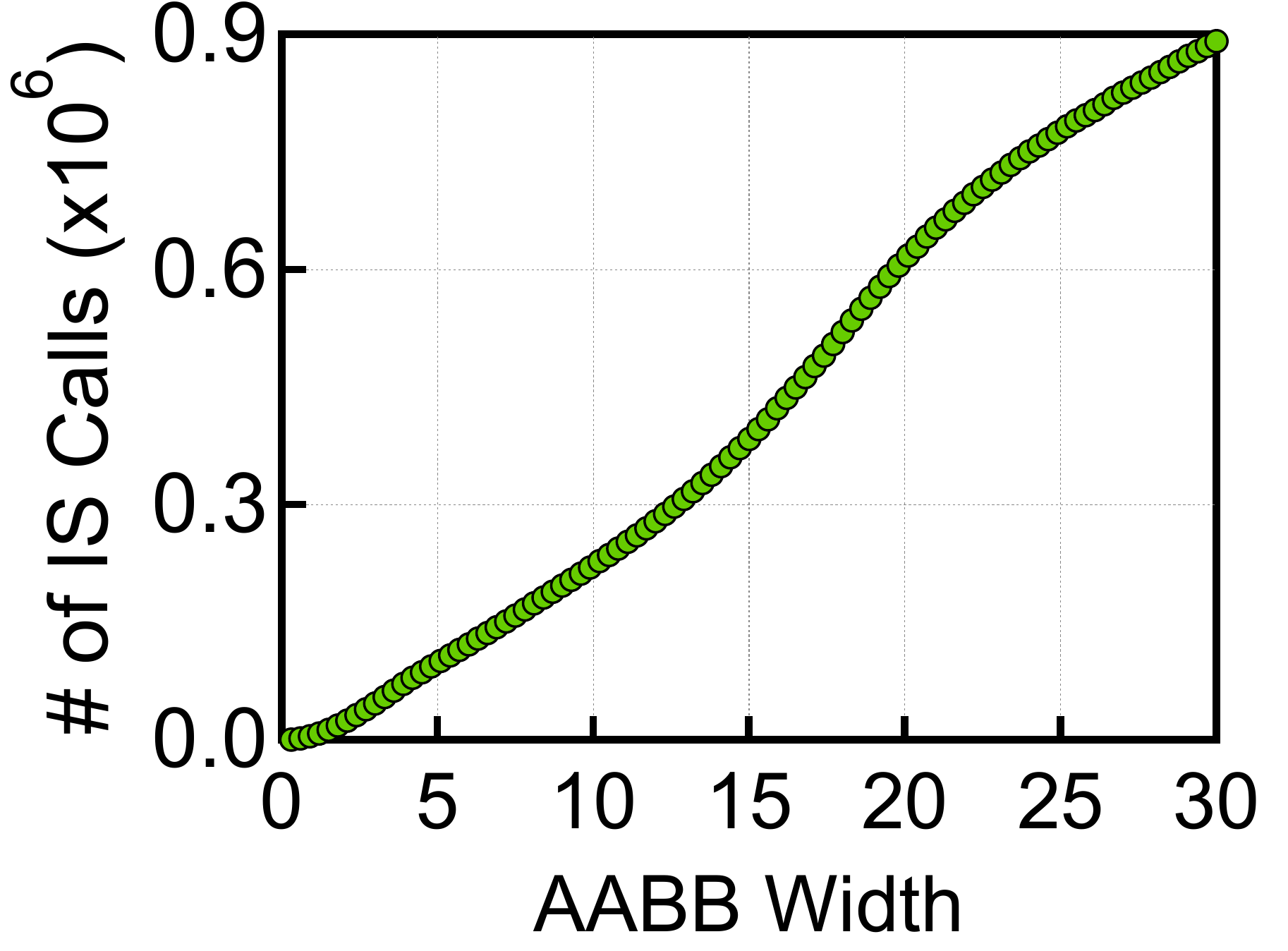}
  \caption{The number of \textsf{IS} calls increases with the width of the AABBs in the BVH. Rays intersect more AABBs when AABBs are larger.}
  \label{fig:is-vs-r}
\end{minipage}
\end{figure*}

\paragraph{Summary} We show the pseudo-code of the range search algorithm in \Lst{code:naiveAlgo}. KNN search is similar except the \is shader would operate a priority queue.

\begin{lstlisting}[language=mylang, numbers=left, aboveskip=\smallskipamount, belowskip=\smallskipamount, mathescape=true, label=code:naiveAlgo, caption={Pseudo-code of range search as ray tracing.}]
input: points, queries, radius, K;

buildBVH(points, radius) {
  foreach point in points
    create an AABB {center=point; width=2*radius};
  return (generate BVH from all AABBs);
}

/* code on the host */
bvh $\leftarrow$ buildBVH(points, radius);
//launch pipeline, starting from the RG shader
traceRays(queries, K, radius, bvh);

/* shader code */
RG_Shader() {
  //OptiX API to get current ray ID;
  rayId $\leftarrow$ optixGetLaunchIndex().x
  ray {origin=queries[rayId], direction=[1,0,0], tmin=0, tmax=1.e-16f};
  //trigger BVH traversal with the ray
  castRay(bvh, ray, count=0);
}

IS_Shader() {
  //called every time a ray intersects a leaf AABB
  ray_origin $\leftarrow$ get the ray origin;
  curPoint $\leftarrow$ get the AABB center;

  if (distance(ray_origin, curPoint) < radius^2) {
    record curPoint as a neighbor;
    if ((count + 1) == K) calls the AH shader;
    else count++;
  }
}

AH_Shader() {
  terminate the current ray;
}
\end{lstlisting}

Lines 3-6 show that we generate an AABB for each point and build the BVH. All the AABBs have the same width (twice the search radius). Lines 17-18 show that each query is mapped to a ray. When the maximum neighbor count $K$ is met, the \is shader calls the \ah shader to terminate the ray.

\subsection{Understanding the Performance}
\label{sec:idea:char}


We characterize key aspects that impact the performance of our algorithm. The performance characterizations point out sources of potential inefficiency and motivate optimizations that we propose in sections that follow. The performance results are obtained from a RTX 2080 Ti GPU. Input data used in this section are from the popular KITTI dataset~\cite{geiger2012we, kitti_raw}. See \Sec{sec:exp} for a complete experimental setup.

\subsubsection{Ray Coherence}
\label{sec:idea:char:rays}

Tree traversal is control-flow intensive. Rays that are spatially distant (``incoherent'' rays in graphics parlance~\cite{pharr1997rendering, aila2010architecture}) will diverge when traversing the BVH. For instance, Ray A and Ray B in \Fig{fig:optix-timeline} exercise different traversal paths and execute different shaders.

OptiX groups every 32 adjacent rays generated in the \rg shader into a warp. This means adjacent rays, if representing spatially-distant queries, will lead to control-flow divergence. Even worse, for tree traversal-based algorithms, control-flow divergences translate to lower memory access efficiency. This is because incoherent rays access different tree nodes as they exercise different traversal paths, increasing the working set size and reducing chances for memory coalescing.

To demonstrate the impact of incoherent rays on neighbor search, we perform a simple experiment, where we assign queries uniformly to the cells in a 3D grid and compare two different query-to-ray mappings: 1) queries are mapped to rays according to the raster-scan order of the grid cells such that adjacent rays represent spatially-close queries, and 2) queries are randomly mapped to rays.

\Fig{fig:random-vs-raster} shows the results. To draw general conclusions, $x$-axis varies the number of queries from 0.27 to 27 millions. Searching with arbitrarily-ordered rays is consistently 5 times slower compared to searching with coherent rays. The performance difference is corroborated by the micro-architectural behaviors. \Fig{fig:random-vs-raster-stat} shows that the search with ordered queries/rays has significantly higher L1/L2 cache hit rate and SM occupancy compared to the incoherent search.

It is worth noting that in the current OptiX implementation, a ray could, at run time (and out of a programmer's hands), be moved to a different thread, warp, or an SM to improve the ray coherency~\cite{optixov}. Our results show that \textit{even with} the run-time coherence optimization by OptiX, performance remains sensitive to the initial query to ray mapping.

\PBox{\textbf{Observation 1}: \textit{Search performance is sensitive to ray coherence, which could be improved by mapping queries to rays such that adjacent rays represent spatially-close queries.}}

\subsubsection{AABB Size.}
\label{sec:idea:char:tra}

The search time strongly correlates with the AABB size in the BVH. To demonstrate the impact of the AABB size, we fix the amount of queries and vary the AABB width in the BVH from 0.3 to 30. \Fig{fig:time-vs-aabb} shows that the search time increases as the AABB width increases.




The reason that the search time correlates with the AABB size is that a larger AABB size means a query is enclosed by more AABBs; thus, the corresponding ray intersects more AABBs. More ray-AABB intersections translate to more BVH traversals on the RT cores (step 1 in the algorithm), which in turn leads to more \is calls on the SMs (step 2 in the algorithm). \Fig{fig:is-vs-r} confirms that the number of \is shader calls increases with the AABB width\footnote{Statistics about the number of traversals are hidden by OptiX. The fact that \Fig{fig:time-vs-aabb} and \Fig{fig:is-vs-r} have the same trend means that the time per \is execution is roughly constant, which we experimentally confirm.}. Interestingly, the number of \is shader calls grows super-linearly. This is because the AABB volume grows \textit{cubicly} w.r.t. its width, so the number of AABBs that a query resides in (i.e., ray intersections) grows cubicly too.

\PBox{\textbf{Observation 2}: \textit{Search time is strongly correlated with the AABB size, which dictates the work on both the SMs and the RT cores. Reducing AABB size reduces the search time.}}

\section{Spatially-Ordered Query Scheduling}
\label{sec:sched}

\begin{figure}[t]
\centering
\includegraphics[width=\columnwidth]{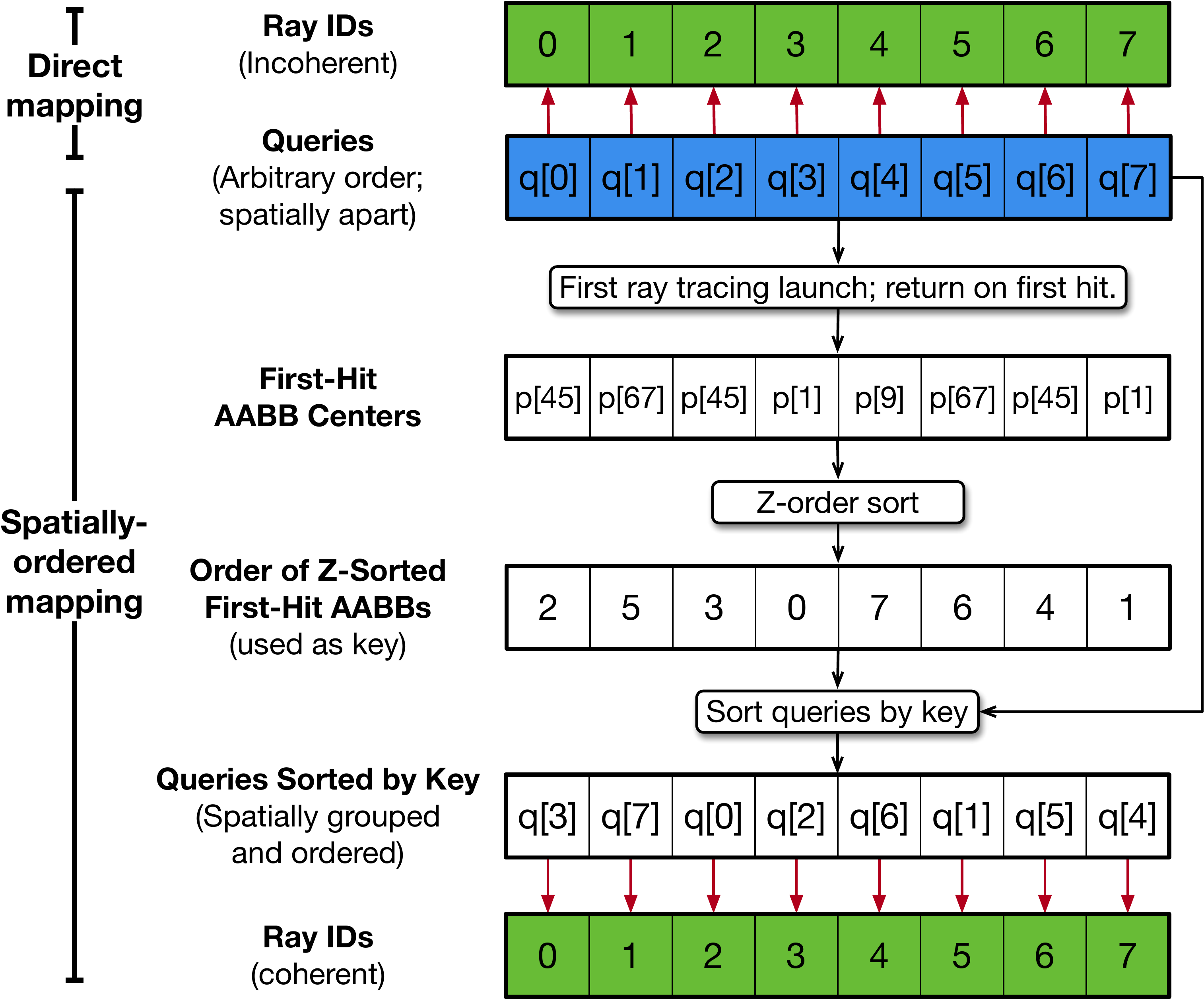}
\caption{Comparison between direct query-to-ray mapping (top) and spatially-order mapping (bottom). \textbf{\textsf{p[]}} denotes the array of points, which are also the AABB centers; \textbf{\textsf{q[]}} denotes the array of queries, which are also the ray origins.}
\label{fig:raysched}
\end{figure}

Each query is mapped to a ray; a direct mapping, shown at the top of \Fig{fig:raysched}, maps queries to rays in the order that the queries appear in the input, which could be arbitrary, leading to incoherent rays. This section introduces a query scheduling technique that tames the ray incoherence and reduces the control-flow divergence in the algorithm.

Our intuition is to group spatially close queries such that adjacent rays follow similar BVH traversal paths. We propose a lightweight grouping algorithm using a simple heuristic: \textit{queries that reside in the same leaf AABB are spatially close} and should be grouped together. In practice, a query is usually enclosed by many leaf AABBs; any such enclosing AABB would provide a useful hint for the query's spatial proximity. That is, we are not interested in a particular enclosing AABB for a query as long as we associate an AABB with each query.


This loose definition of spatial proximity allows us to group queries very efficiently --- as another ray tracing problem! In particular, finding an enclosing AABB for a query can be done by casting a ray for the query and immediately terminating the ray once the \textit{first} \is shader is called, essentially returning the first intersecting leaf AABB of each query. This ray tracing is very efficient because it invokes the \is shader only once for each ray without traversing the entire BVH.

\begin{lstlisting}[language=mylang, numbers=left, aboveskip=\smallskipamount, belowskip=\smallskipamount, mathescape=true, label=code:reorder, caption={Pseudo-code of neighbor search with ray reordering, which is done through an initial ray tracing that terminates when the first leaf AABB is found for each ray.}]
bvh$\leftarrow$buildBVH(points, radius);
// initial search with K = 1
FirstHitAABBs$\leftarrow$traceRays(queries, 1, radius, bvh);
reorderQueries(queries, FirstHitAABBs);
// second search with the actual K
traceRays(queries, K, radius, bvh);
\end{lstlisting}

\Lst{code:reorder} shows the pseudo-code of the search algorithm with query grouping. Essentially we perform ray tracing twice: the first time is with a maximum neighbor count $K = 1$ (Line 3) and the second time is with the actual search $K$ (Line 6). After the first search, all the queries have a first-hit AABB ID; queries with the same ID are then grouped together.

One issue remains: how should different groups be ordered? Recall that each leaf AABB represents a point in the search space. Thus, the order of the first-hit AABBs is essentially the order in which their corresponding search points appear in the input, which could be arbitrary.

To introduce order into the first-hit AABBs, we simply sort their corresponding points (i.e., AABB centers) in a Morton (Z) order. This is done by the function {\color{editorGreen}{\texttt{reorderQueries()}}} in \Lst{code:reorder}, and is implemented in a CUDA kernel, which operates on the first-hit AABB data produced by the shaders directly in the device memory without extra memory copies. \Fig{fig:raysched} compares the the spatially-ordered query-to-ray mapping (bottom) with the direct mapping (top).

\section{Query Partitioning and Bundling}
\label{sec:part}

This section introduces a technique to suppress BVH traversals. The idea is to partition queries and build a specialized BVH for each partition with the smallest possible AABB size without violating correctness. We first describe the basic idea (\Sec{sec:part:idea}), followed by an algorithm to determine the (near-)optimal partitioning (\Sec{sec:part:opt}).



\subsection{Query Partitioning}
\label{sec:part:idea}

\begin{figure*}[t]
    \centering
    \subfloat[Calculating the megacell of a query by incrementally growing from the cell that contains the query. The growth stops when either the sphere boundary is reached or at least $K$ neighbors are found.]{
      \label{fig:megacell}
      \includegraphics[height=1.4in]{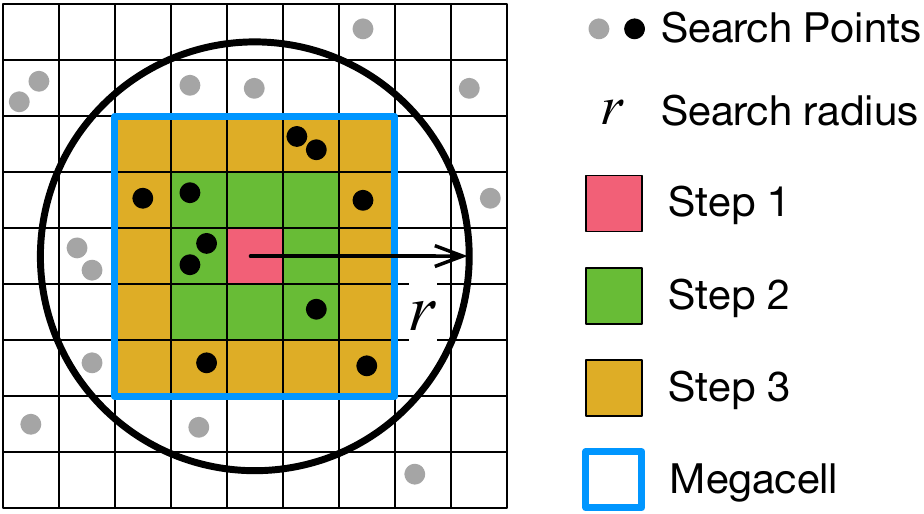}
    }
    \hspace{10pt}
    \centering
    \subfloat[$P1$ is outside of $Q$'s megacell that contains 3 points, but is among the 3 nearest neighbors of $Q$ (as $d1 < d2$).]{
      \label{fig:knn-cornercase}
      \includegraphics[height=1.4in]{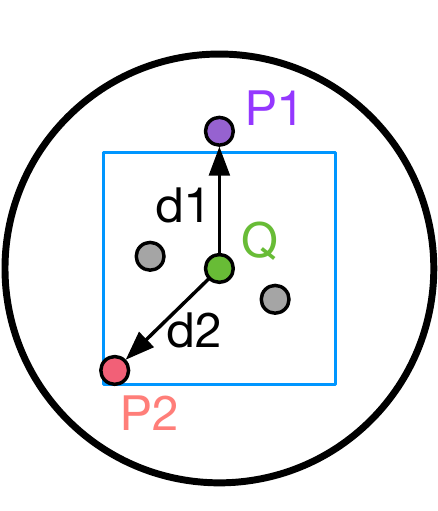}
    }
    \hspace{10pt}
    \subfloat[For KNN search, the AABB must circumscribe the sphere that circumscribes the megacell. The AABB width is $\sqrt{2}a$ for 2D search (illustrated here) and $\sqrt{3}a$ for 3D search (as the AABB is 3D and the circumcircle becomes the circumsphere), where $a$ is the megacell width.]{
      \label{fig:knn-partition}
      \includegraphics[height=1.4in]{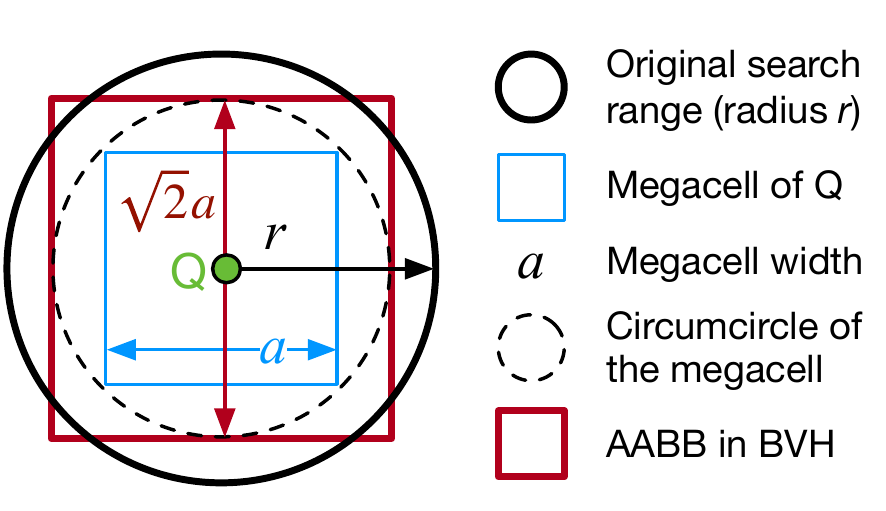}
    }
    \caption{Determining the megacell and AABB size used for neighbor search.}
    \label{fig:knn-aabb}
\end{figure*}

In our baseline algorithm, all the queries share the same BVH. We observe, in \Sec{sec:idea:char:tra}, that the search time is strongly correlated with the AABB size in the BVH, because larger AABBs result in more traversals and \is shader calls, which increases the search time.

Our idea is to, for each query, identify an AABB size that is \textit{just large enough} to ensure correctness. In this way, instead of using one monolithic BVH for all queries, queries are partitioned into different partitions, each with a unique BVH that minimizes the amount of search work for that partition.

In theory, the AABB width must be twice as the search radius $r$ provided by users (see \Fig{fig:rnn-test}). However, it is possible to use smaller AABBs for a query if its $K$ neighbors can be found within a smaller radius. We use an iterative method over a uniform grid to identify the proper AABB size for a query. This is illustrated in \Fig{fig:megacell}.

We first create a uniform grid over the scene containing all the search points. For each query, we then calculate the least amount of grid cells that contain $K$ neighbors. This calculation is done by starting from the cell that contains the query, and iteratively growing the cells along all six directions (or four in the case of 2D search). The growth stops just before the $r$-radius sphere boundary is reached or at least $K$ neighbors are found. We call the final collection of cells a \textit{megacell}. Naturally, the largest possible megacell is the cube (square) that is inscribed by the sphere (circle).

\paragraph{Determining AABB Size} Given the megacell of a query, the next step is to decide the AABB size used to build the BVH, which differs between range search and KNN search.

In range search, the AABB size can be safely set to the megacell size. The actual search would simply return the $K$ neighbors from the megacell, which are guaranteed to be within the distance $r$ from the query $Q$. An additional benefit now is that the \is shader does not have to perform the sphere test anymore (Step 2 in \Sec{sec:idea:basic}), since any query that is enclosed by the AABB is guaranteed to be enclosed by the sphere. This leads to significant performance gains.

The situation is slightly more complicated for KNN search. Even if a megacell contains $K$ neighbors of $Q$, it does not mean the \textit{nearest} $K$ neighbors of $Q$ are in the megacell. \Fig{fig:knn-cornercase} shows a counter-example where $P1$, which is just outside of the megacell, is part of the 3 nearest neighbors of $Q$ even though the megacell contains 3 neighbors.

We \textit{can}, however, guarantee that the circumscribed sphere of the megacell will contain the $K$ nearest neighbors of $Q$. This is illustrated in \Fig{fig:knn-partition}. To guarantee correctness, a conservative estimation would be to set the AABB width $w = \sqrt{2}a$ for 2D search (as illustrated in \Fig{fig:knn-partition}) and $w = \sqrt{3}a$ for 3D search, where $a$ is the megacell width.

We use a simple heuristic to reduce $w$: assuming that the point density is locally uniform within and around a megacell, a sphere with the same volume as the megacell should contain $K$ neighbors of $Q$. Thus, we use a $w=2\sqrt[3]{3/(4\pi)}a$ (in 3D search)\footnote{The megacell volume is $a^3$, and the sphere volume is $\frac{4}{3}\pi (\frac{w}{2}) ^3$, where $a$ is the megacell width and $w$ is the sphere diameter (i.e., AABB size) to be solved for. Thus, $w$ is $2\sqrt[3]{3/(4\pi)}a$ to ensure equi-volume.}. We find this heuristics to be sufficient (for correctness) from the datasets we evaluate (\Sec{sec:exp}). One could further relax $w$ if an application is amenable to approximate neighbor search, which we discuss in \Sec{sec:conc}.


\paragraph{Algorithm Summary} \Lst{code:part} shows the pseudo-code of the search algorithm with query partitioning. Lines 1-5 calculate the megacell size for each query. In the end, the queries are naturally split into different partitions, each with a unique AABB size. Lines 7-10 generate a BVH for each partition; each partition is then searched separately using the corresponding BVH. Queries in each partition could be further spatially-ordered as described in \Sec{sec:sched}.

\begin{lstlisting}[language=mylang, numbers=left, aboveskip=\smallskipamount, belowskip=\smallskipamount, mathescape=true, label=code:part, caption={Pseudo-code of neighbor search with query partitioning. Each partition has a different BVH.}, escapeinside={(*@}{@*)}]
grid $\leftarrow$ generate grid from search points;
foreach query in queries
  megaCellWidth $\leftarrow$ gen megacell for query (*@in@*) grid;
  AABBSize $\leftarrow$ megaCellWidth$\sqrt{2}$; // $\sqrt{3}$ for 3D search
  partitions[AABBSize].add(query);

foreach p in partitions
  bvh $\leftarrow$ buildBVH(points, p.AABBSize/2);
  queries = partitions[p];
  traceRays(queries, K, radius, bvh);
\end{lstlisting}


We implement the megacell calculation in CUDA. An important parameter is the grid resolution. Intuitively, small grid cells lead to a more accurate megacell estimation because the stride of each growth step is smaller, but also increase the memory consumption. In our implementation, we use the smallest cell size allowed by the GPU memory capacity.

\subsection{Bundling the Partitions}
\label{sec:part:opt}

By default, each partition is launched separately with its corresponding BVH (as shown in \Lst{code:part}). However, this strategy might be sub-optimal. This is because each partition requires constructing a unique BVH. In cases where the search time saving is smaller than the BVH construction overhead (e.g., when the partitions are small), generating many partitions degrades performance.

We propose an algorithm that optimally bundles partitions together to minimize the overall search time. The idea is to first generate as many partitions as described before, and then decide how to combine partitions together by analytically modeling the cost of bundling partitions. The overall ideas for KNN search and range search are the same, but differ in details. We focus on KNN search here, and leave the details of range search to Supplementary Material~\ref{sec:supp:rangecost}.

\paragraph{Cost Model} As a first-order approximation, the total search cost $T$ is the sum of the cost of each of the $P$ partitions, which in turn is the sum of the BVH construction cost ($T_{build}$) and the actual search cost ($T_{search}$):
\begin{align}
	T = \sum_{i=0}^{P}{(T_{build}^{i} + T_{search}^{i})}.
\end{align}

While Nvidia discloses little detail about their BVH construction algorithm and implementation, we empirically find that the BVH construction time is linearly correlated with the number of AABBs in the BVH (see details in Supplementary Material~\ref{sec:supp:bvh}). We model the BVH construction time as linearly scaling with the number of AABBs $M$:
\begin{align}
	T_{build} = k_1 M.
\end{align}

The search cost for a partition is dictated by the number of \is shader calls, which is a product of the number of queries ($N$) and the number of \is calls per query. The number of \is calls a query makes is equivalent to the number of leaf AABBs that the query resides in, which in turn is the product of the AABB volume and the point density (i.e., average number of points per unit volume). Therefore, the search time $T_{search}$ is modeled as:
\begin{align}
	T_{search} = k_2 N \rho S^3,
\end{align}

\noindent where $S$ is the AABB width and $\rho$ is the point density. Since each partition's megacell, by construction, contains just about $K$ points, $\rho$ can be estimated by $K/C^3$, where $C$ is the megacell width of the partition.

While there are two unknown coefficients $k_1$ and $k_2$ in our modeling, knowing their ratio is sufficient to compare the \textit{relative} costs of partitioning strategies. This ratio can be obtained offline through profiling the BVH construction time per AABB and the \is shader execution time per call. On RTX 2080, this ratio is about 1:15000. Absent the offline profiling, we fall back to the default strategy (\Lst{code:part}).

\paragraph{Bundling Algorithm} Given the cost modeling, let us explain why there exists an optimal bundling. The crux is that bundling partitions increases the total search cost but reduces the BVH construction cost.

Specifically, when we bundle a partition $P_i$ (with $N_i$ queries, an AABB width of $S_i$, a point density $\rho_i$) with another partition $P_j$ (with $N_j$ queries, an AABB width of $S_j$, a point density $\rho_j$), we eliminate one unit of BVH construction cost; meanwhile, the combined partition will have an AABB width of $max(S_i, S_j)$. Thus, the search cost of this new partition is greater than the individual search cost of $P_i$ and $P_j$ combined (assuming the point density does not change abruptly):
\begin{align}
  k_2 (N_i \rho_i + N_j \rho_j) [max(S_i, S_j)]^3 > k_2 (N_i \rho_i S_i^3 + N_j \rho_j S_j^3).
\end{align}


The goal of our bundling algorithm is to determine how to optimally bundle the available partitions to minimize the total cost. In theory, this is a combinatorial optimization, which in general is intractable. Fortunately, we empirically observe that this problem has a special structure that lends itself to be solved efficiently at run time. We leave the proof to Supplementary Material~\ref{sec:supp:bundle}, and state the conclusion below.

To find the optimal bundling, we first sort all partitions in the ascending order of their query counts; we then start from the last partition and linearly scan toward the first partition. At each stop, we bundle all the partitions that have been scanned, leave the rest unbundled, and calculate the total cost. The bundling strategy that has the lowest cost $T$ wins.

\section{Evaluation}
\label{sec:eval}

After describing our evaluation methodology (\Sec{sec:exp}), we show the overall speedup of \proj (\Sec{sec:eval:perf}), followed by teasing apart the contributions of different optimizations (\Sec{sec:eval:opt}). Finally, we study the sensitivity of \proj with respect to search configurations (\Sec{sec:eval:sen}).

\begin{figure}[t]
    \centering
    \subfloat[Speedups on RTX 2080.]{
      \label{fig:2080res}
      \includegraphics[width=\columnwidth]{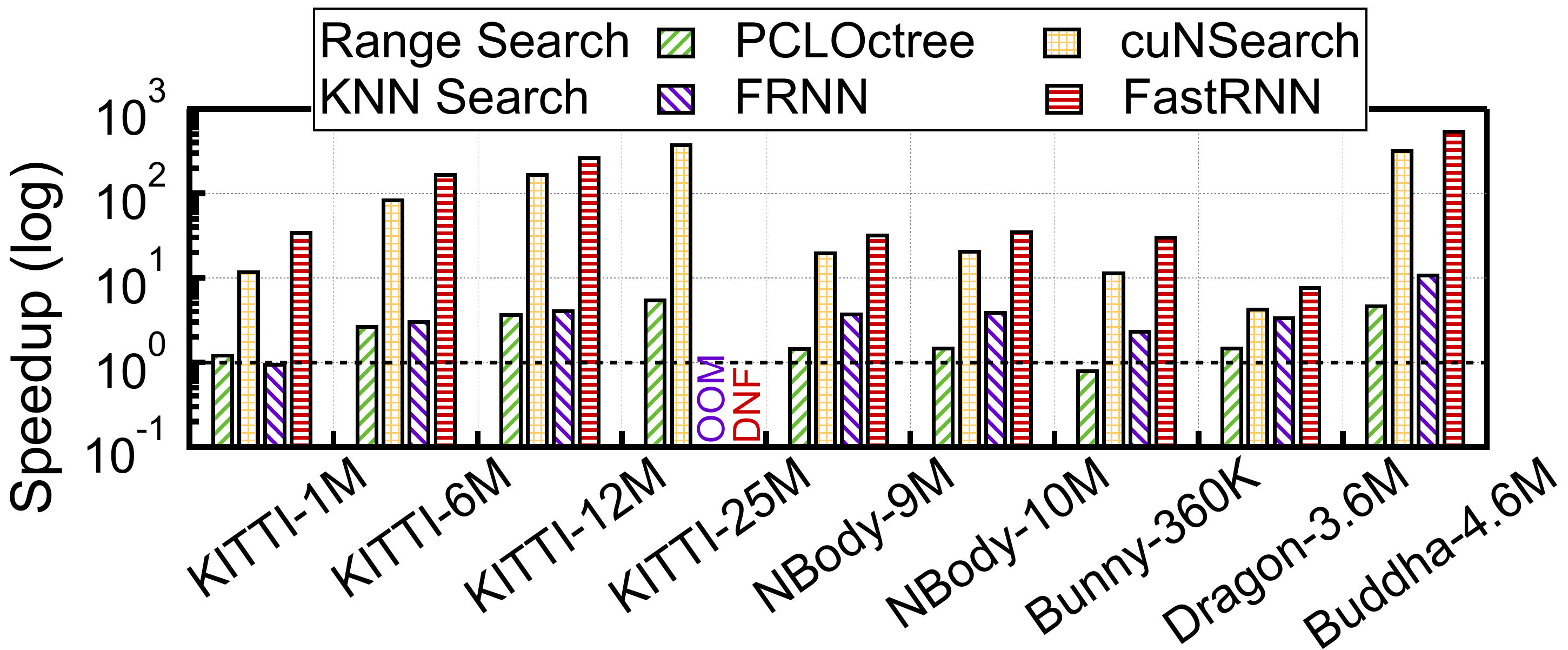}
    }
    \\
    \subfloat[Speedups on RTX 2080Ti.]{
      \label{fig:2080tires}
      \includegraphics[width=\columnwidth]{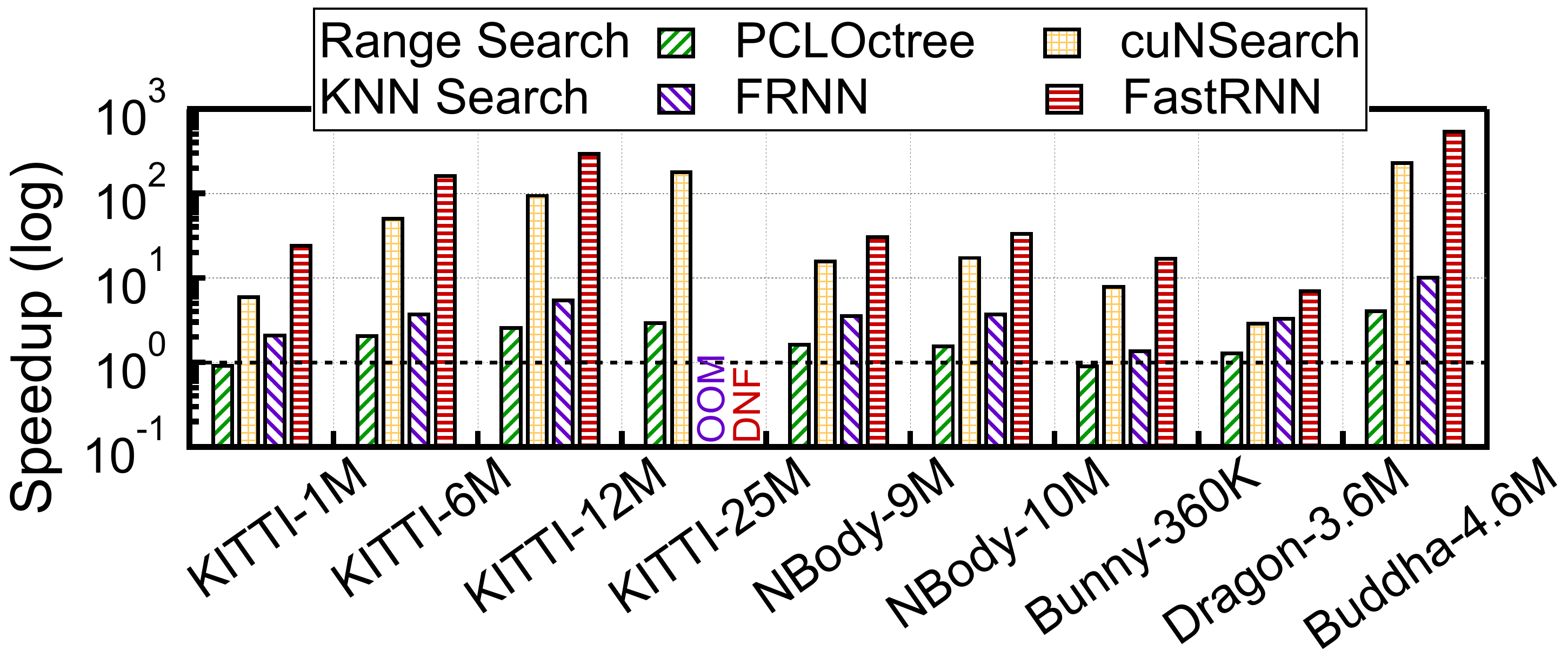}
    }
    \caption{Speedup of \proj over the baselines (log-scale). OOM denotes that the baseline ran out of memory; DNF denotes that the baseline did not finish within the time that would have given \proj a 1,000 speedup.}
    \label{fig:res}
\end{figure}

\subsection{Evaluation Methodology}
\label{sec:exp}

\paragraph{Environment} We implement our algorithm in OptiX 7.1; the entire program is compiled with nvcc V11.4.48. We evaluate the performance on two Turing GPUs: a RTX 2080Ti (68 RT cores, 4352 CUDA cores, 11 GB GDDR6) and a RTX 2080 (46 RT cores, 2944 CUDA cores, 8 GB GDDR6).


\paragraph{Baselines} We compare with four GPU baselines, which are built and evaluated in the same environment as \proj.

\begin{itemize}
	\item \textbf{\textsc{cuNSearch}}~\cite{hoetzlein2014fast, cunsearchcode} is an optimized CUDA library used in many scientific computing applications such as the widely popular SPlisHSPlasH fluid simulator~\cite{splishsplashcode}. cuNSearch has only a range search implementation.
	\item \textbf{\textsc{FRNN}}~\cite{frnn} is a drop-in replacement for (and about 10 times faster than) the KNN search in PyTorch~\cite{pytorch3dknn}. We instrument the code to measure just the CUDA time without the Python wrapper overhead.
	\item \textbf{\textsc{PCLOctree}} is an octree-based CUDA implementation in Point Cloud Library (PCL)~\cite{Rusu_ICRA2011_PCL, pclcode}, a widely-used library for computational geometry, graphics, and vision. PCLOctree is available for both KNN search and range search ($K$ must be 1 for KNN search).
	\item \textbf{\textsc{FastRNN}}~\cite{evangelou2021fast} is a recent work that leverages the RT cores for KNN search only and \textit{without} the various optimizations that we propose in this paper.
\end{itemize}

\paragraph{Why These Baselines?} Both \textsc{cuNSearch} and \textsc{FRNN} are grid-based algorithms. The performance gains of \proj over them highlight the benefits of using a tree structure (i.e., BVH) with hardware acceleration.

Similar to \proj, \textsc{PCLOctree} also uses a hierarchical data structure, i.e., the Octree, which is a space-partitioning structure rather than an object-partitioning structure like the BVH. Comparing with \textsc{PCLOctree} shows the benefits of hardware-supported object partitioning. Comparing to \textsc{FastRNN} quantifies the optimizations proposed in this paper.


\paragraph{Datasets} We use three datasets covering three  domains where neighbor search is critical. We first use the LiDAR-generated point clouds from the KITTI self-driving car dataset, which is commonly used in computer vision and robotics research~\cite{geiger2012we, kitti_raw}. To evaluate the scalability, we combine point cloud frames to obtain three final frames with a point count of 1M, 12M, and 25M, respectively.

The second dataset consists of three 3D-scanned models from the Stanford 3D Scanning Repository~\cite{3dscanrep}: Bunny (360K points), Asian Dragon (3.6M points), and Buddha~\cite{curless1996volumetric} (4.6M points), all of which are widely used for graphics research. Finally, we use a cosmological N-body simulation dataset~\cite{millenniumdata, springel2005simulating} from the Millennium Simulation Project~\cite{millennium}. The dataset has two traces with 9M and 10M particles (galaxies) each.

Apart from covering three representative application domains where neighbor search is critical, these three datasets also allow us to evaluate \proj under different point distributions. Points in the KITTI self-driving car dataset are mostly distributed in the $xy$-plane (the ground) and while being confined in a very narrow $z$-range (height). Points in the other two datasets occupy the entire 3D space, but point distribution in the cosmological simulation is much more \textit{non-uniformly} than that in the 3D scanning dataset. Points in cosmological simulation represent galaxies in the universe; galaxy distribution is naturally non-uniform\footnote{On scales of order 1 to 10 Mpc/h, the galaxy distribution is roughly hierarchical clustering (fractal), where 1 Mpc/h is of order the spacing between galaxies. On scales much larger than 10 Mpc/h the matter distribution very slowly approaches uniformity\no{, with very large-scale structures on scales of 100 Mpc/h (superclusters)}. The Millennium Simulation dataset runs 500 Mpc/h on a side and, thus, exhibits the non-uniform distribution.}.


\subsection{Overall Performance Analysis}
\label{sec:eval:perf}

On RTX 2080, \proj provides a (geomean) 2.2$\times$ and 44.0$\times$ speedup over \textsc{PCLOctree} and \textsc{cuNSearch}, respectively, on range search, and provides a (geomean) 3.5$\times$ and 65.0$\times$ speedup over \textsc{FRNN} and \textsc{FastRNN}, respectively, on KNN search. \Fig{fig:2080res} shows the per-input speedup.

We observe that: 1) the speedup increases when the number of points increases, and 2) the speedup on range search is generally lower than that on KNN search. Let us elaborate. 

\paragraph{Across Input Scales} On the two smallest inputs (KITTI-1M and Bunny-360K), \proj has only limited speedup (up to 2$\times$) over the fastest baselines \textsc{PCLOctree} and \textsc{FRNN}. The speedups on larger inputs (e.g., KITTI-12M and Buddha-4.6M) are at least 5--10$\times$. \Fig{fig:2080tires} shows the results on RTX 2080Ti; the same trend holds.

\begin{figure}[t]
    \centering
    \subfloat[KNN search time distribution.]{
      \label{fig:pie2080knn}
      \includegraphics[width=.475\columnwidth]{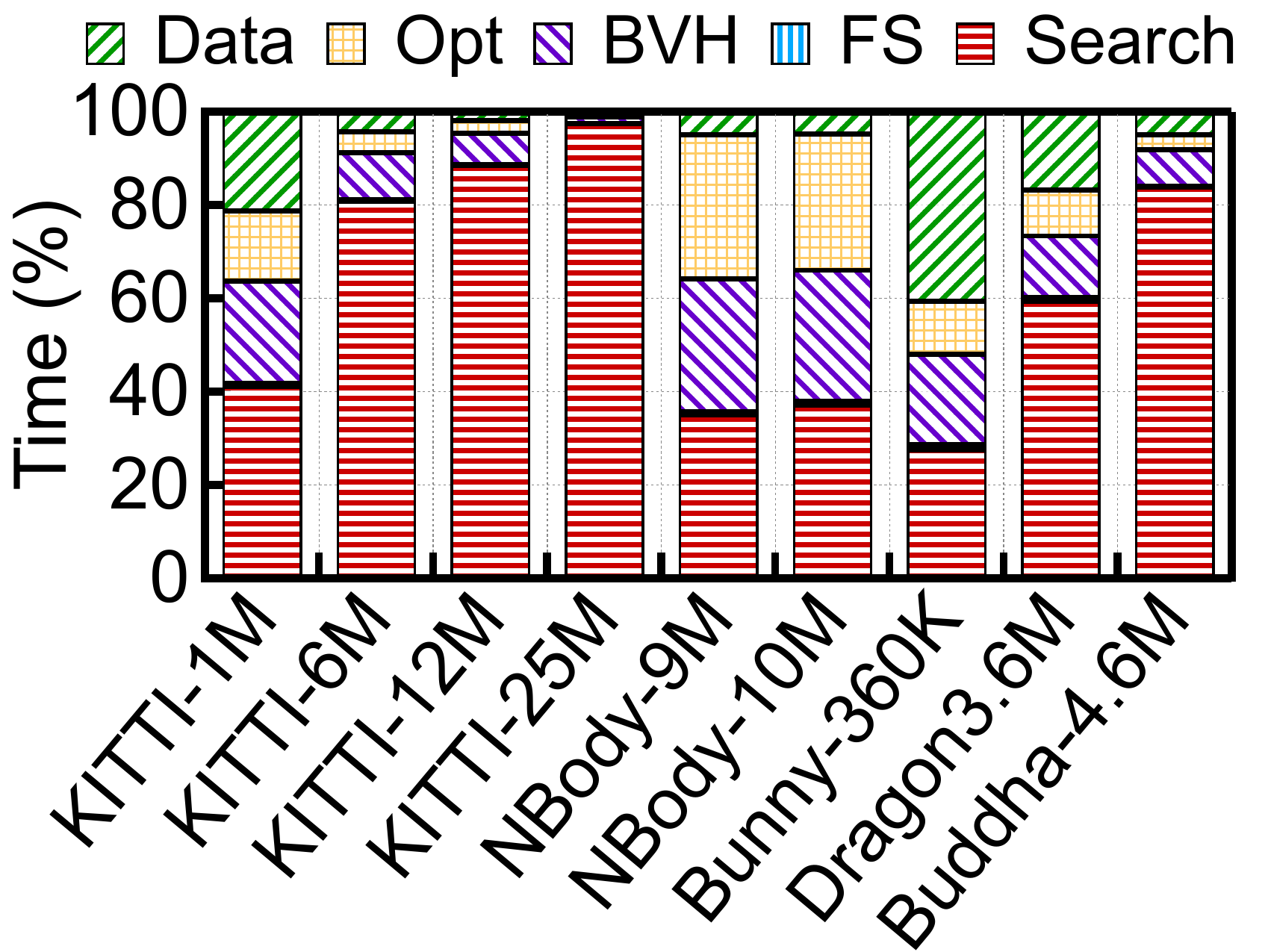}
    }
    \hfill
    \subfloat[Range search time distribution.]{
      \label{fig:pie2080radius}
      \includegraphics[width=.475\columnwidth]{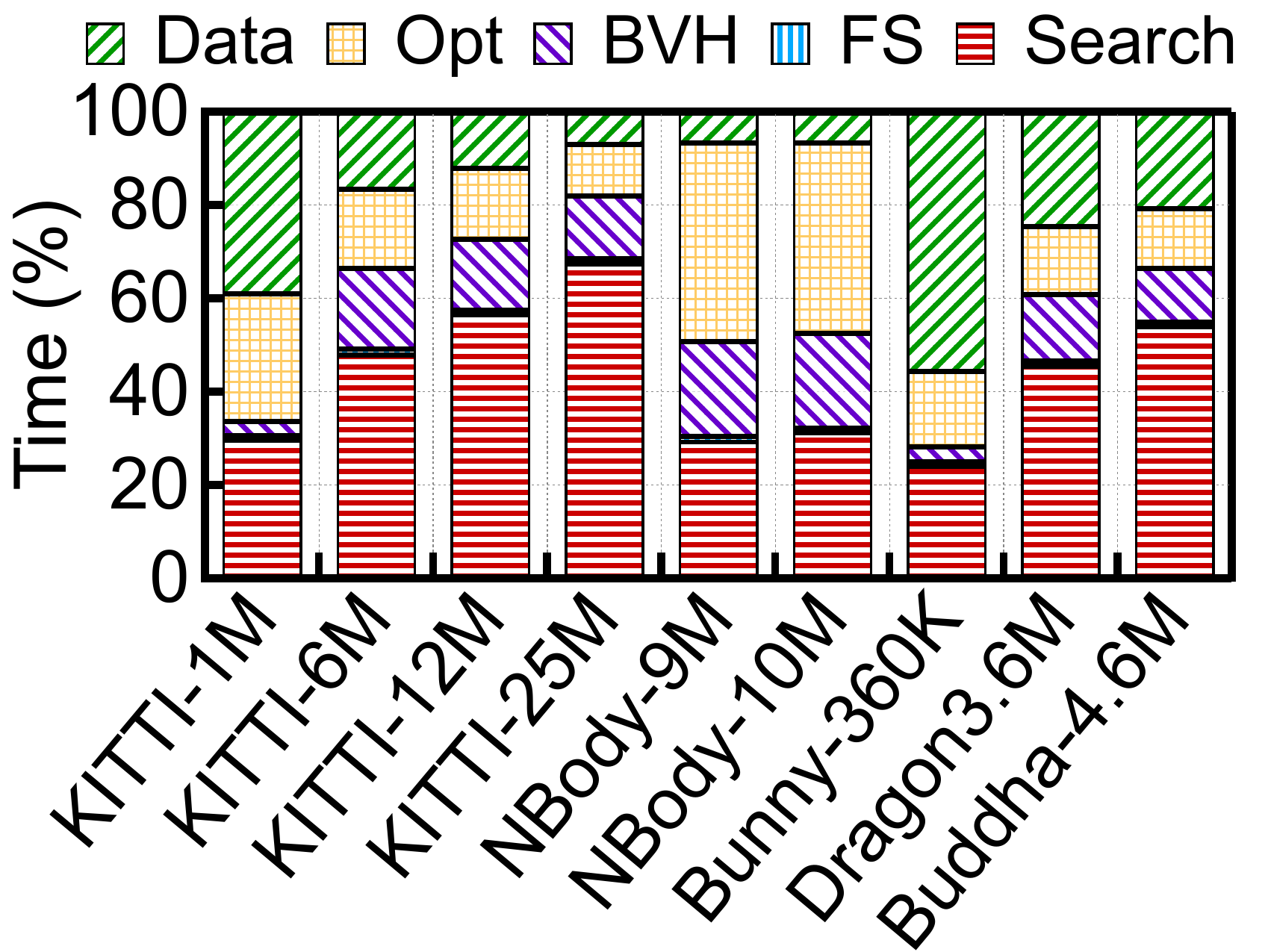}
    }
    \caption{Time distribution on RTX 2080.}
    \label{fig:pie}
\end{figure}

To understand why the speedup changes with the scale of the input, \Fig{fig:pie2080knn} breaks down the KNN search time on RTX 2080 into five components: data transfer time (\texttt{Data})\footnote{This includes times to copy data to and from the device memory. The former is not hidden, but the latter is almost completely hidden.}, the overhead of applying optimizations (\texttt{Opt}), including reordering and partitioning queries, time spent on generating the BVHs (\texttt{BVH}), the first search to find the first-hit AABBs for queries (\texttt{FS}), and the second (actual) search (\texttt{Search}). The execution times of smaller inputs (e.g., KITTI-1M and Bunny-360K) are dominated by non-search related tasks, diminishing the gains from accelerating the search.

The two N-body inputs are interesting cases: they have large numbers of points but still spend more than half of the time on non-search related tasks. A close examination shows that the spatial density of their points (i.e., galaxies used in the N-body simulation) varies a lot. Thus, queries have different megacell sizes and fall into different partitions. As a result, \proj spends much time generating the partitions (\texttt{Opt}) and building the different BVHs (\texttt{BVH}).


\paragraph{KNN vs. Range Search} We also observe that the speedup on KNN search is generally higher than that on range search. This is because KNN search spends more than in the actual search than range search due to the need to manipulate a priority queue. This time distribution difference is evident by comparing \Fig{fig:pie2080knn} and \Fig{fig:pie2080radius}, which show the time distribution for KNN search and range search, respectively. For instance, on KITTI-12M \proj spends 88.5\% of the time on the actual search (\texttt{Search}) under KNN search, which decreases to only 63.5\% under range search.

\subsection{Teasing Apart Optimizations}
\label{sec:eval:opt}

\begin{figure}[t]
    \centering
    \subfloat[KITTI-12M.]{
      \label{fig:optsen-kitti12m-2080}
      \includegraphics[width=.475\columnwidth]{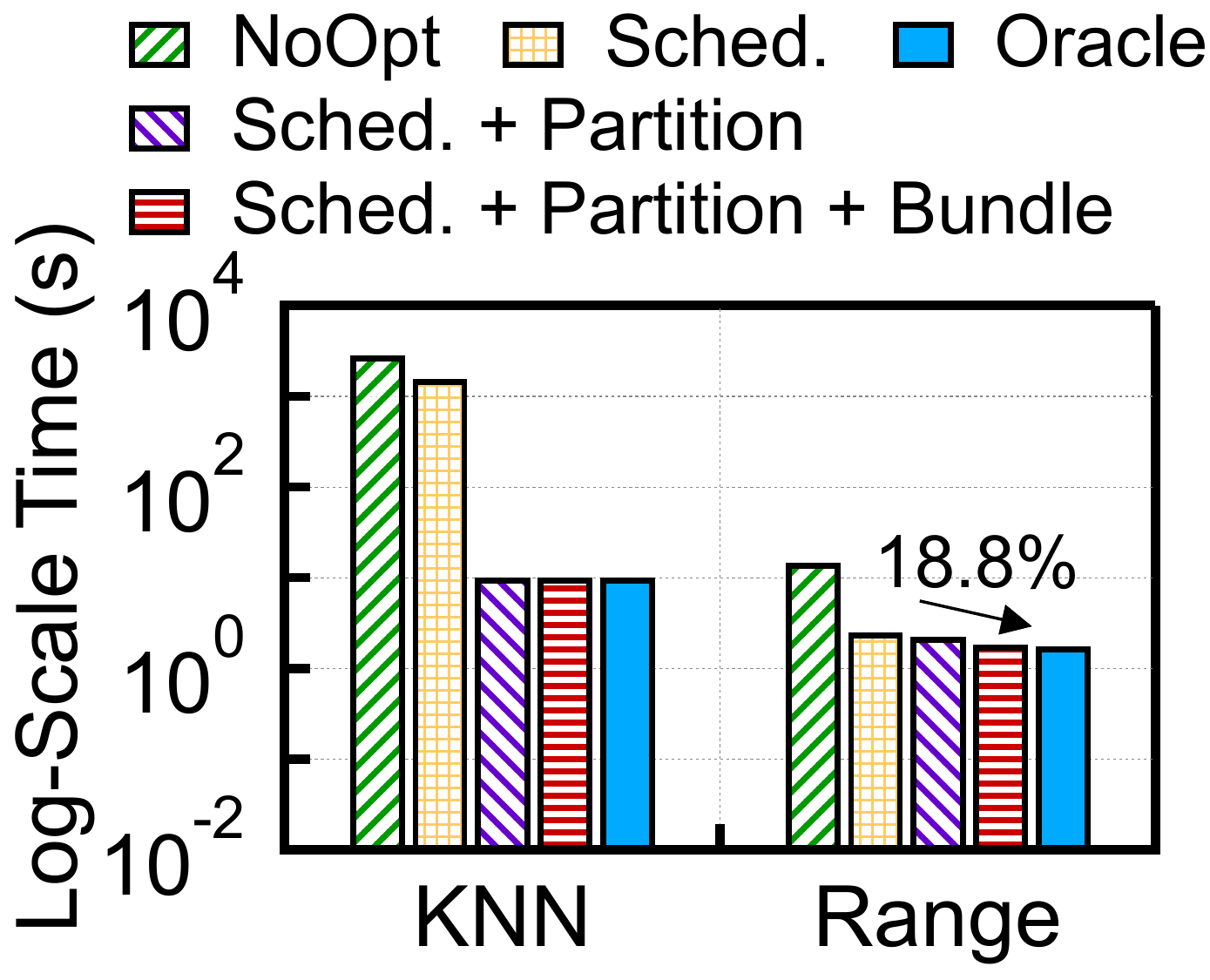}
    }
    \hfill
    \subfloat[NBody-9M.]{
      \label{fig:optsen-nbody9m-2080}
      \includegraphics[width=.475\columnwidth]{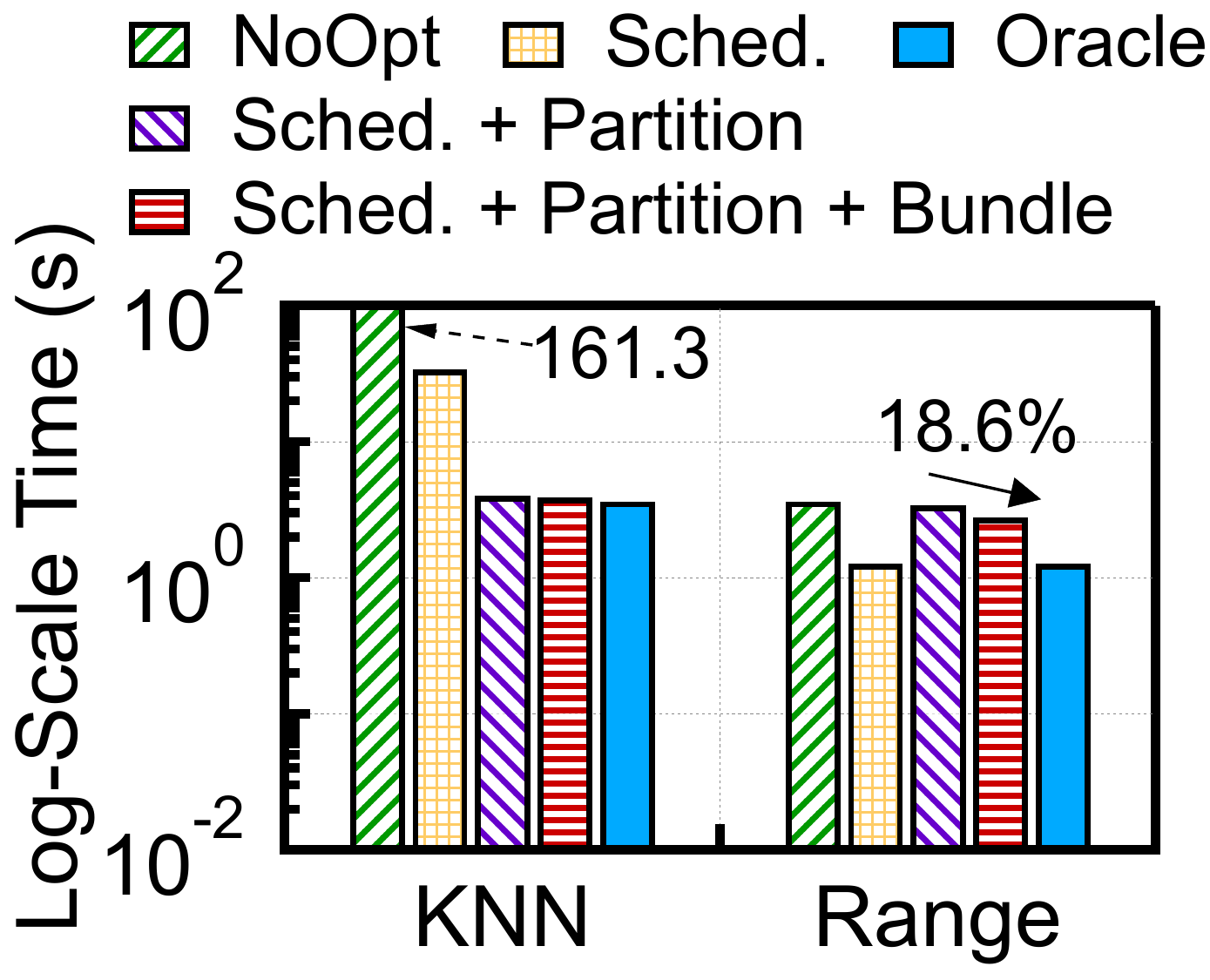}
    }
    \caption{Effects of our optimizations on RTX 2080.}
    \label{fig:optsen}
\end{figure}

The optimizations we propose in this paper are critical to the performance benefits of \proj. Using two representative inputs, \Fig{fig:optsen} compares the performance of five variants of our algorithm on RTX 2080 (KITTI-12M in \Fig{fig:optsen-kitti12m-2080} and NBody-9M in \Fig{fig:optsen-nbody9m-2080}):

\begin{itemize}
	\item \texttt{NoOpt}: no optimization;
	\item \texttt{Sched.}: query scheduling only (\Sec{sec:sched});
	\item \texttt{Sched. + Partition}: query scheduling and query partitioning (\Sec{sec:part:idea});
	\item \texttt{Sched. + Partition + Bundle}: query scheduling, partitioning, and bundling (\Sec{sec:part:opt});
	\item \texttt{Oracle}: assuming a priori knowledge of 1) whether to partition, and 2) the best bundling strategy through an offline exhaustive search (infeasible for run time).
\end{itemize}

\paragraph{Scheduling} Comparing to \texttt{NoOpt}, ray scheduling improves the performance by 1.8$\times$ and 5.9$\times$ on KNN and range search, respectively, for KITTI-12M; the speedups are 4.7$\times$ and 2.9$\times$ for NBody-9M. The speedups not only come from better ray coherence but also because the overhead of finding the first-hit AABBs is negligible, which is evident in \Fig{fig:pie}, where the \texttt{FS} category is virtually invisible.

\paragraph{Partitioning} Query partitioning is much more effective for KNN search. On KITTI-12M, partitioning provides a 154.4$\times$ and 1.1$\times$ speedup for KNN and range search, respectively, on top of ray scheduling.

Recall that query partitioning improves speed by suppressing tree traversals and \is calls. KNN search needs many more traversals (as it must find the $K$ \textit{nearest} neighbors whereas range search terminates tree traversal whenever $K$ neighbors are found), and the cost of an \is shader call in KNN search is (3--6$\times$) higher than that in range search. Thus, query partitioning is more effective to KNN search.

Interestingly, query partitioning degrades performance on NBody-9M. As discussed early, the points in N-body simulations are non-uniformly distributed, which results in a high partitioning and BVH construction overhead (\Fig{fig:pie}).

\paragraph{Bundling} Bundling provides an additional 18.8\% and 18.6\% performance gain on range search for the two inputs, but has little impact on KNN search. This is because KNN search, with its hefty costs of \is shader and traversal, typically uses all the partitions anyways (i.e., no bundling), which our bundling algorithm accurately captures.

Overall, our bundling algorithm is effective, leading to a performance that is within 3\% of \texttt{Oracle} for KITTI-12M. The \texttt{Oracle} for NBody-9M is achieved when partitioning is disabled, whereas \proj always assumes partitioning. A future improvement to \proj is to estimate the points' spatial density before deciding whether to partition.

\subsection{Sensitivity Analysis}
\label{sec:eval:sen}

\begin{figure}[t]
    \centering
    \subfloat[Sensitivity of speedup to $r$.]{
      \label{fig:2080_buddha_r}
      \includegraphics[height=1.1in]{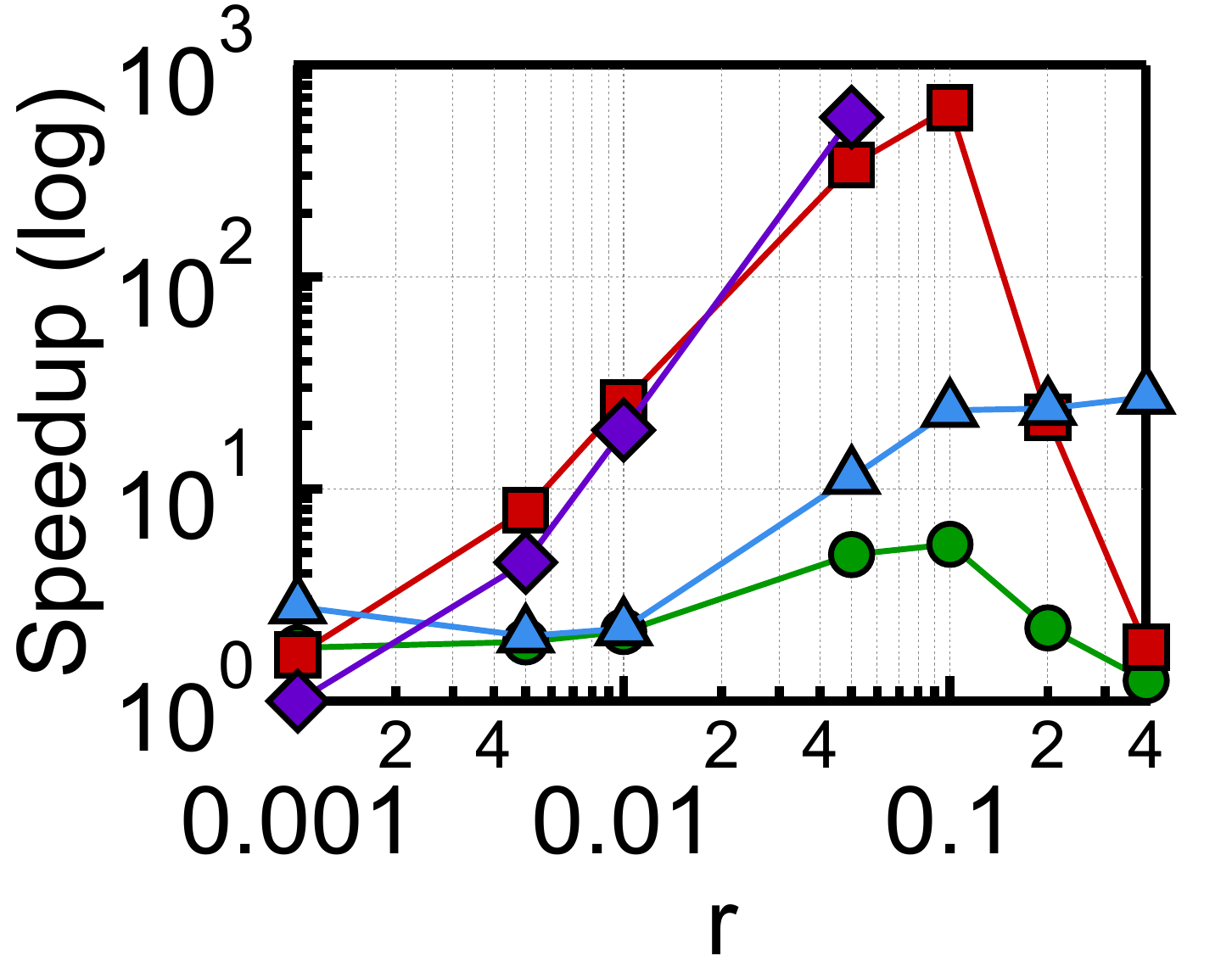}
    }
    \subfloat[Sensitivity of speedup to $K$.]{
      \label{fig:2080_buddha_k}
      \includegraphics[height=1.1in]{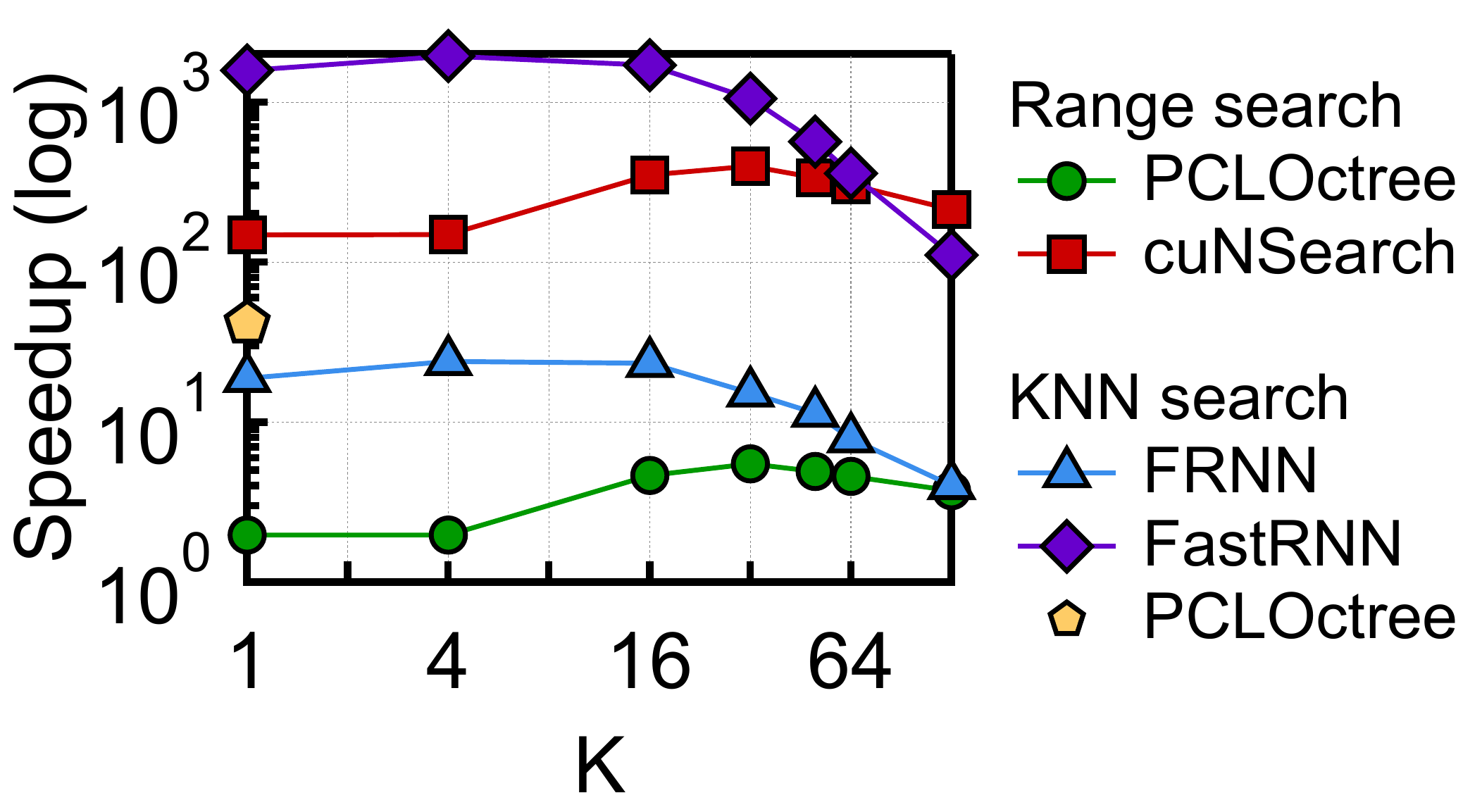}
    }
    \caption{Range search speedup of Buddha-4.6M on RTX 2080 as $r$ and $K$ change. \textsc{PCLOctree} supports only K=1 for KNN search, so it has only one KNN data point. The missing data points of \textsc{FastRNN} is because it did not finish within the time that would have given \proj a 1,000 speedup.}
    \label{fig:sen}
\end{figure}

\proj offers speedups across a range of $r$ and $K$ values. Using range search on Buddha-4.6M as an example, \Fig{fig:2080_buddha_r} and \Fig{fig:2080_buddha_k} show how the speedup on RTX 2080 varies with $r$ and $K$, respectively.

\paragraph{Sensitivity to $r$} As $r$ increases our speedups increase initially, because a larger $r$ means a larger AABB and, thus, more search work that can be accelerated. For range search, however, the speedups (over \textsc{PCLOctree} and \textsc{cuNSearch}) decrease (still >1) as the search radius $r$ exceeds 0.1. This is because the points in Buddha are bounded in a $1^3$ cube; the search sphere under, say, $r=0.4$, covers almost the entire cube. Queries can thus quickly find neighbors under a large $r$, so the search terminates quickly, in which case the overhead of \proj (e.g., building BVH, ray scheduling) are more pronounced, leading to lower speedups.




\paragraph{Sensitivity to $K$} As $K$ grows, \proj's speedup generally increases, because a larger $K$ leads to more search work that can be accelerated by \proj. The speedup degrades when $K$ becomes too big (e.g., 128). We find that this is because the bundling algorithm tends to be overly aggressive under a larger $K$. We leave it to future work to investigate a better bundling algorithm under large $K$s.

\section{Related Work}
\label{sec:related}

\Sec{sec:bg:ns} provides an overview of neighbor search. \Sec{sec:exp} discusses neighbor search algorithms in low-dimensional space. This section focuses on work related to ray tracing.

\paragraph{RTX Beyond Ray Tracing} Recent papers have started using (Nvidia) ray tracing hardware to accelerate workloads beyond ray tracing, including both rendering workloads~\cite{zellmann2020high, morrical2019efficient, zellmann2020accelerating} and non-rendering workloads~\cite{wald2019rtx, evangelou2021fast, salmon2019exploiting}.

Among them, Evangelou et al.~\cite{evangelou2021fast} and Zellmann et al.~\cite{zellmann2020accelerating} are the closest to our paper; the former uses RT cores for 3D KNN search and the latter uses RT cores for 2D range search. This paper provides a unified neighbor search algorithm with two generally-applicable optimization that significantly improves the search performance. In addition, we provide a detailed performance characterization of using ray tracing for neighbor search (\Sec{sec:idea:char}).

\paragraph{Ray Incoherence} Literature is rich with techniques that improve ray tracing performance, much of which is focused on taming incoherent rays~\cite{eisenacher2013sorted, pharr1997rendering, aila2010architecture, shkurko2017dual, gribble2008coherent, barringer2014dynamic}, which lead to both branch divergences and memory inefficiencies.

All existing techniques target rendering. We propose an efficient reordering technique specialized to neighbor search, which exposes two opportunities. First, all the rays (queries) are known at the beginning, enabling a global reordering; in contrast, incoherent rays in rendering are dynamically generated (e.g., bounces) and, thus, are usually locally reordered. In this sense, our ray scheduling is similar to wavefront ray tracing~\cite{laine2013megakernels, shkurko2017dual} but with only one wavefront. Second, our rays have the same direction, so reordering involves only the ray origin (3D) whereas reordering for rendering usually involves both ray origin and direction (6D).



\section{Conclusion and Future Work}
\label{sec:conc}

Conventional GPUs, while initially built for rasterization-based graphics, are now widely used as general-purpose accelerators for parallel algorithms. Emerging ray tracing-based GPUs beg the question: can we use the ray tracing hardware for workloads beyond ray tracing? This paper uses neighbor search as a case study and provides a positive answer. We show that effectively exploiting the ray tracing hardware requires carefully mapping work items (i.e., queries in our case) to rays and suppressing excessive tree traversals. The analyses and techniques in this paper provide useful insights in broadening the utility of ray tracing hardware.

\paragraph{Approximate Neighbor Search} Many applications do not require exact neighbor search. \proj is amenable to approximation, sometimes with quantitative error bounds. We discuss two opportunities here. First, in building a BVH for a query (or a query partition) one could use an AABB size smaller than what is strictly required. Using a smaller AABB would reduce the number of neighbors returned but also provide performance gains, since performance is sensitive to AABB size as established in \Sec{sec:idea:char:tra}.

Second, one could elide Step 2 in the search algorithm (\Sec{sec:idea:basic}), i.e., treating any query that resides in an AABB as residing in the inscribed sphere. Under this approximation, given a query range $r$ all the returned neighbors are bound to be within a distance $\sqrt{3}r$ of the query. Speedups from this approximation would be significant, given that Step 2 is much more costly than Step 1.

\paragraph{General-Purpose Irregular Processor} Given the success of today's GPGPUs for regular algorithms, one would naturally wonder how a ray tracing-based GPU can be used as a general-purpose processor for \textit{irregular} applications. Ray casting is fundamentally a tree traversal problem, which is central to many irregular applications beyond graphics.

Realizing this vision requires us to carefully rethink the architecture, run-time system, and programming model~\cite{graphicsblog}. Hardware-wise, Nvidia's RT cores are specialized for BVH traversal with a specific branching logic (bounded intersection checking), which is not easily extensible to more generic traversals. Meanwhile, the run-time ray scheduler, a performance-critical component, should ideally be customized to a particular algorithm. Our paper shows that the default OptiX scheduler optimized for ray tracing is sub-optimal for neighbor search. Finally, it would be interesting to explore programming models that free programmers from constantly thinking about rays and geometry.

\section{Acknowledgements}
\label{sec:ack}

The author expresses gratitude to Prof. Kelly Douglass and Prof. Segev BenZvi, both of Department of Physics and Astronomy at University of Rochester, for pointing to the N-body simulation dataset, and to Prof. Michael Vogeley of Department of Physics at Drexel University for explaining the point density distribution in N-body simulation. The author also thanks Tiancheng Xu and Boyuan Tian for providing help in setting up PCL. The work was supported, in part, by NSF under grants \#2044963 and \#2126642.
\section*{Appendix}
\label{sec:supp}

\renewcommand{\thesubsection}{\Alph{subsection}}

\subsection{Cost Model for Range Search}
\label{sec:supp:rangecost}

The BVH construction cost $T_{build}$ of range search is the same as that of KNN search. The search cost $T_{search}$ is different from that in KNN search. In particular, the number of \is shader calls per query in range search is always $K$, as the search terminates as soon as $K$ neighbors are found. That is:
\begin{align}
	T_{search} = k_3 N K,
\end{align}

\noindent where $S$ is the AABB width, $N$ is the number of queries in a partition. $k_3$ is the time of an \is shader in range search.

$k_3$ depends on the megacell size of a partition. When the megacell size does not touch the search sphere, the \is shader can skip the ray-sphere intersection test, because the corresponding query, being inside the megacell, is guaranteed to reside in the sphere. When the megacell size touches the sphere, the \is shader has to perform the ray-sphere intersection test, because a query residing in the megacell does not guarantee that the query will reside in the sphere. As a result, the ratio of $k_1$ to $k_3$ varies. On RTX 2080, the ratio is about 20:1 in the former case and is 2:1 in the latter case.

\subsection{Modeling BVH Construction Time}
\label{sec:supp:bvh}

\Fig{fig:gasbuildtime} shows how the BVH time varies with the number of AABBs. We regress a linear fit for the correlation with an $R^2$ of 0.996, indicating a strong linear relationship.

\subsection{Derivation of the Optimal Bundling Algorithm}
\label{sec:supp:bundle}

\begin{figure}[t]
\centering
\begin{minipage}[t]{0.48\columnwidth}
  \centering
  \includegraphics[trim=0 0 0 0, clip, height=1.3in]{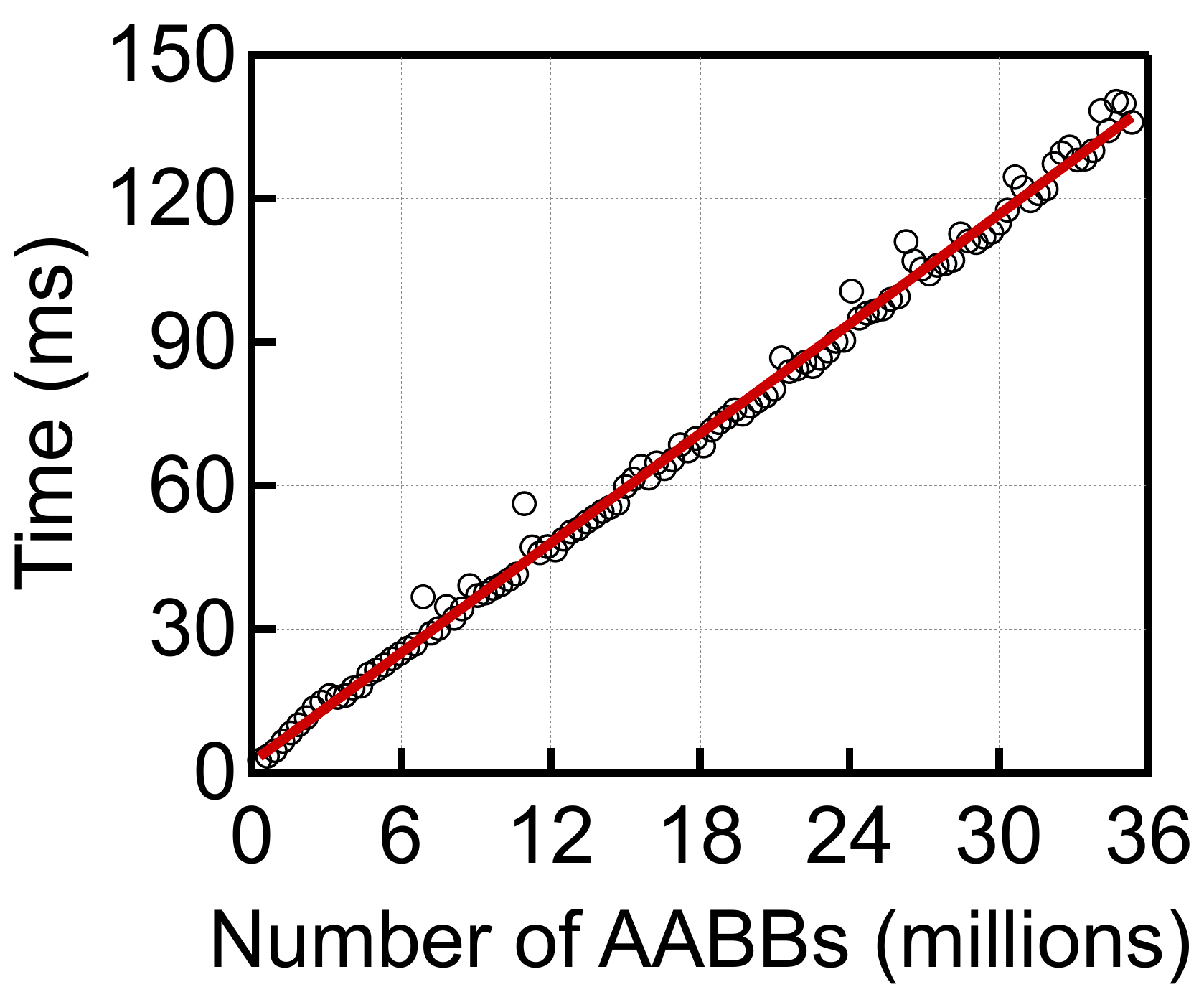}
  \caption{BVH construction time is linearly correlated with the number of AABBs.}
  \label{fig:gasbuildtime}
\end{minipage}
\hfill
\begin{minipage}[t]{0.48\columnwidth}
  \centering
  \includegraphics[trim=0 0 0 0, clip, height=1.3in]{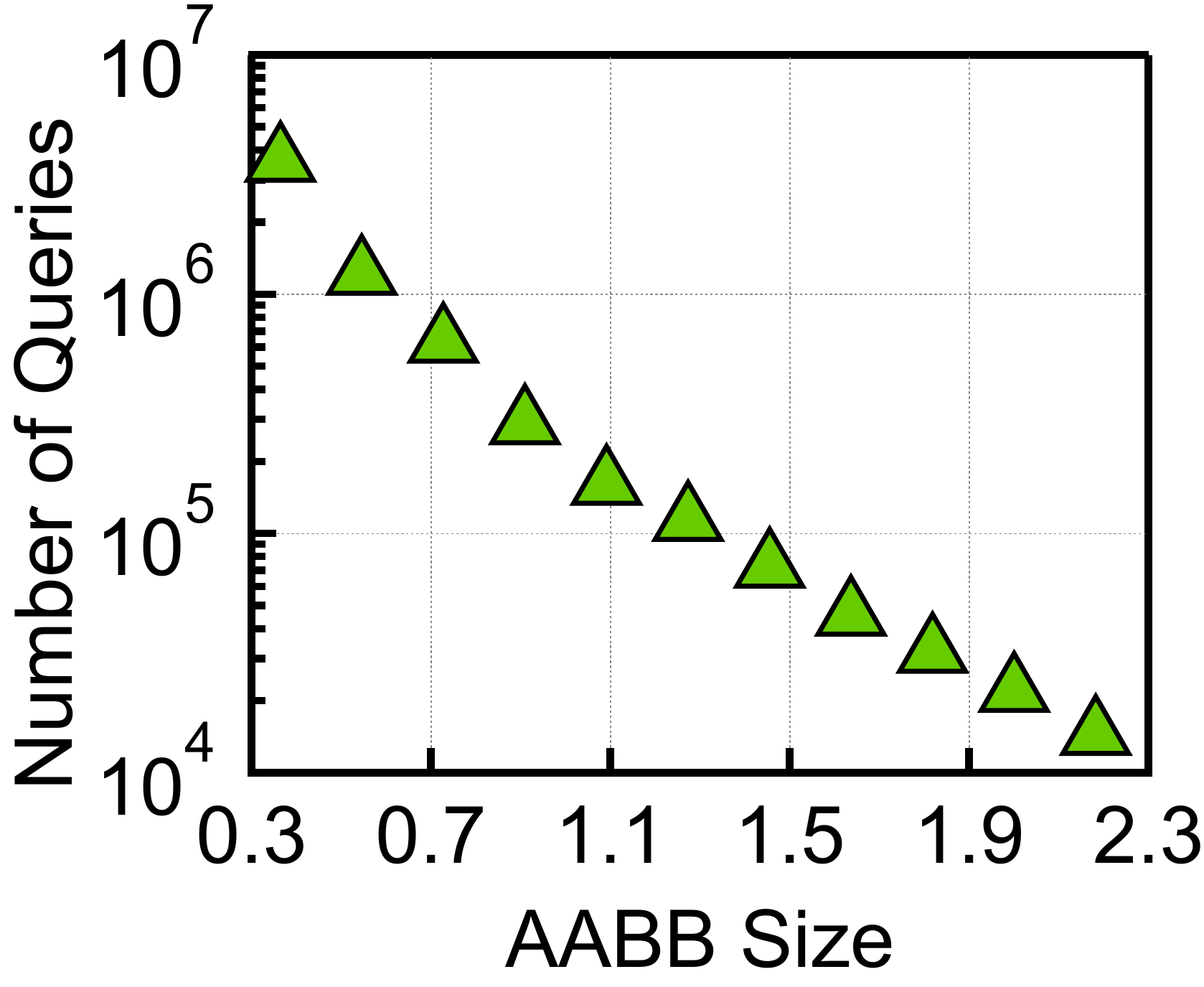}
  \caption{Number of queries and the AABB size in a partition are inversely correlated.}
  \label{fig:queryhist}
\end{minipage}
\end{figure}

We find that as the AABB size of a partition increases the number of queries in the partition usually decreases. \Fig{fig:queryhist} shows a typical query distribution over the AABB size when searching about 6 millions queries in total. This makes statistical sense, because usually only a handful of sparsely located queries need a large AABB, whereas most of queries should be captured by small AABBs.

\begin{figure}[t]
\centering
\includegraphics[width=\columnwidth]{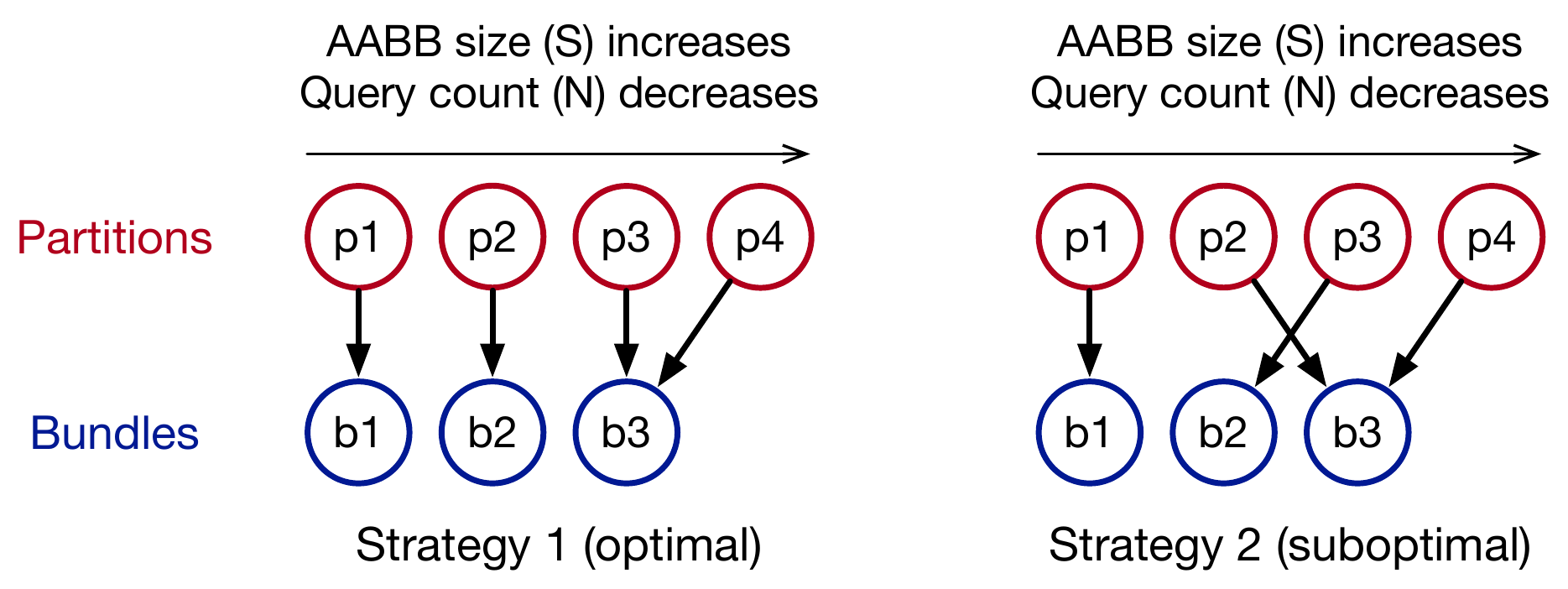}
\caption{Partition bundling. Under the empirical observation that the AABB size and the query count of the partitions are inversely correlated, bundling strategy 1 is optimal, whereas strategy 2 is sub-optimal.}
\label{fig:optpart}
\end{figure}

With this observation, our bundling algorithm is divided into two steps. First, we prove the following theorem:

\textbf{\underline{Theorem}}: \textit{if} the optimal number of bundles is $M_o$ ($1 \leq M_o \leq M$, where $M$ is the number of available partitions), the optimal bundling strategy is one where the $(M_o-1)$ partitions that have the most queries are \textit{not} bundled and the remaining partitions are combined into one bundle.

\Fig{fig:optpart} illustrates it using a simple example, which has four partitions that are sorted according to the ascending order of their AABB sizes $S$, which is also the descending order of their query counts $N$ (based on the empirical observation above). Assuming that the optimal number of bundles is three, the optimal bundling strategy is strategy 1 (left figure), where the first two partitions have their own bundles and the last two partitions are combined. Its search cost is (omitting the constant coefficient $k_2$):
\begin{align}
	T_{search}^{(1)} = N_1 \rho_1 S_1^3 + N_2 \rho_2 S_2^3 + (N_3\rho_3+N_4\rho_4) S_4^3.
\end{align}

We can prove this by contradiction. Without losing generality, let us assume that strategy 2 in \Fig{fig:optpart}, where $p2$ and $p4$ are combined, is optimal. The resultant search cost is thus (again omitting the coefficient $k_2$):
\begin{align}
	T_{search}^{(2)} = N_1\rho_1 S_1^3 + N_3\rho_3 S_3^3 + (N_2\rho_2+N_4\rho_4) S_4^3.
\end{align}

Since $S_2 < S_3 < S_4$, $S_2 = \sqrt{3} C_2$, $S_3 = \sqrt{3} C_3$ we have:
\begin{align}
	\rho_2 = K / C_2^3 > K / C_3^3 = \rho_3.
\end{align}

Combined with $N_2 > N_3$, we have:
\begin{align}
	T_{search}^{(1)} - T_{search}^{(2)} = N_3\rho_3(S_4^3 - S_3^3) - N_2\rho_2(S_4^3 - S_2^3) < 0.
\end{align}

Since both strategies require three BVH constructions (as there are three bundles), strategy 2 has a higher total cost ($T_{build} + T_{search}$) than that of strategy 1, contradicting the proposition that strategy 2 is optimal.

The theorem essentially allows us to find the optimal bundling in constant time \textit{given} an $M_o$. The problem then is reduced to finding the optimal $M_o$, which is a linear-time problem: we linearly search all the possible $M_o$ values; for each $M_o$ value ($1 \leq M_o \leq M$), we estimate its cost according to the bundling strategy given by the theorem (constant time) and pick the $M_o$ that has the lowest search cost.

\bibliographystyle{ACM-Reference-Format}
\bibliography{refs}

\renewcommand{\thesubsection}{\Alph{subsection}}
\setcounter{figure}{0}
\setcounter{equation}{0}
\setcounter{page}{1}

\end{document}